\documentclass[preprint,12pt]{elsarticle}
\usepackage{setspace}
\usepackage{titlesec}
\usepackage{caption}
\titlespacing*{\chapter}{0pt}{4ex plus 1ex minus .2ex}{3ex plus .2ex}
\titlespacing*{\section}{0pt}{3ex plus 1ex minus .2ex}{2ex plus .2ex}
\titlespacing*{\subsection}{0pt}{3ex plus 1ex minus .2ex}{2ex plus .2ex}
\titlespacing*{\subsubsection}{0pt}{3ex plus 1ex minus .2ex}{2ex plus .2ex}

\usepackage[english]{babel}
\usepackage[letterpaper,top=2cm,bottom=2cm,left=3cm,right=3cm]{geometry}

\usepackage{amsmath,amssymb,mathtools,graphicx,booktabs,multirow}
\usepackage[colorlinks=true,allcolors=blue]{hyperref} 

\journal{Marine Structures} %

\begin{document}
\begin{frontmatter}

\title{A hybrid global–local computational framework for ship hull structural analysis using homogenized model and graph neural network}

\author[1]{Yuecheng Cai}
\author[1,2]{Jasmin Jelovica\corref{cor1}}
\cortext[cor1]{Corresponding author}
\ead{jasmin.jelovica@ubc.ca}

\address[1]{Department of Mechanical Engineering, The University of British Columbia, Vancouver, BC V6T 1Z4, Canada}
\address[2]{Department of Civil Engineering, The University of British Columbia, Vancouver, BC V6T 1Z4, Canada}

\begin{abstract}
This study presents a computational framework for global–local structural analysis of ship hull girders that integrates an equivalent single-layer (ESL) model with a graph neural network (GNN). A coarse-mesh homogenized ESL model efficiently predicts the global displacement field, from which degrees of freedom (DOFs) along stiffened panel boundaries are extracted. A global-to-local DOF mapping and reconstruction procedure is developed to recover detailed boundary kinematics for local analysis. The reconstructed DOFs, together with panel geometry and loading, serve as inputs to a heterogeneous graph transformer (HGT), a subtype of GNN, which rapidly and accurately predicts the detailed stress and displacement fields for any panel within the hull girder. The HGT is trained using high-fidelity 3D panel finite element model with reconstructed boundary conditions, enabling it to generalize across varying panel geometries, loadings, and boundary behaviors. Once trained, the framework requires only the global ESL solution in order to generate detailed local responses, making it highly suitable for optimization. Validation on three box beam case studies demonstrates that the global prediction error is governed by the coarse-mesh ESL solution, while the HGT maintains high local accuracy and clearly outperforms conventional ESL-based stress estimation method.
\end{abstract}


\end{frontmatter}

\section{Introduction}\label{sec1}

The development of efficient ship structures is essential for advancing the maritime industry, enhancing economic performance, operational safety, environmental sustainability, and technological innovation. Achieving these improvements demands rigorous structural analysis from early stages of the design. Effective surrogate models are required, given that analysis needs to be performed repeatedly in order to optimize the structural performance. 

At the early design stages, naval architects rely primarily on classification society rules to define high-level structural characteristics. These prescriptive rules provide a fast way to generate preliminary designs but may lack accuracy and fail to account for the unique characteristics of each vessel. Consequently, inherent inaccuracies in the initial structural design can propagate into later stages of the design process. On the other hand, finite element analysis (FEA) can provide high-fidelity information on structural performance, requiring typically six to eight shell elements between stiffeners to capture stress, displacement, and buckling behavior with acceptable accuracy \cite{clough1990original}. Global structural FEA can then easily comprise of millions of degrees of freedom, which is computationally intensive to solve, especially if material nonlinearity is involved. This is especially a problem if the structure is to be optimized properly. State-of-the-art global optimization algorithms can require 100,000 design evaluations to converge on ship hull girder problems \cite{jelovica2024improved, cai2023neural}. Therefore, there is a critical need for fast and accurate tools for structural analysis of ships during the early stages of development. 

\subsection{Equivalent single layer and homogenization methods for ship structural analysis}\label{sec1_1}

To efficiently evaluate the global response of ship structures, coarse mesh global finite element analysis (FEA) is commonly employed, necessitating the use of homogenization methods. These methods simplify complex structures by representing them with equivalent homogeneous properties, capturing essential mechanical behavior while significantly reducing computational demand \cite{sanchez1980non, milton2002theory}. Conventional techniques, such as the idealized structural unit method (ISUM) \cite{yukio1984idealized} and Smith's method \cite{smith1977influence}, divide structures into discrete superelements to efficiently predict ultimate strength and collapse modes. Although widely adopted in classification society rules \cite{iacs2014common}, these methods primarily focus on global-scale failure prediction, lacking the capability to provide local stress information for comprehensive global analysis, nor can they account for interaction between deformation modes and loadings in different directions.

The equivalent single layer (ESL) approach further advances homogenization by representing some of the missing stiffness terms in the formulations above, allowing for more accurate analysis \cite{noor1996computational, reddy2003mechanics}. Hughes \cite{hughes2005ship} developed the foundational work in the field of ship structures, which integrates stiffeners into plate elements as orthotropic properties, enabling efficient coarse mesh global analysis. Subsequent refinements by Kumar and Mukhopadhyay \cite{kumar2000finite} included bending effects using discrete Kirchhoff-Mindlin elements, although limitations in shear stress representation remained. Further improvements by Avi et al. \cite{avi2015equivalent} and Gonçalves et al. \cite{goncalves2016homogenization} enhanced ESL modeling by incorporating membrane-bending coupling, shear stiffness and nonlinear stiffness updates.

Additional research utilizing ESL and homogenization methods for hull girder structures can be found in \cite{putranto2021ultimate,putranto2022ultimate,korgesaar2023equivalent}, where the buckling response and ultimate strength of stiffened panels are investigated. Recently, this method has been extended to simulate an entire ship by Putranto et al. \cite{putranto2022application}, demonstrating the capability of ESL-based homogenization approaches in capturing global ship responses efficiently.

While ESL effectively predicts load-displacement behavior of a hull girder and the averaged response of stiffened panels, it lacks the capability to quantify the detailed stress and displacement fields in the structures since the information of local geometry is lost through homogenization. This necessitates additional sub-modeling steps \cite{thompson2017ansys, ma2012finite, samanta2009fatigue}, which significantly increases the complexity. To overcome this difficulty, one option is to integrate ESL with advanced deep learning-based techniques for local stress and displacement prediction. Such integration would enable accurate prediction of complex local responses based on global analysis, effectively bridging the gap between global efficiency and local accuracy in ship structural analysis.

\subsection{Surrogate models for structural analysis}\label{sec1_2}

Surrogate models have become essential in engineering to reduce computational costs associated with complex simulations, particularly for repetitive analyses required in optimization and design. Traditional surrogate methods such as multivariate adaptive regression splines (MARS), kriging (KRG), radial basis functions (RBF), and the response surface method (RSM) are well-established; however, they often struggle with geometrical complexity and nonlinearity inherent in advanced structural systems \cite{chen2006review}. Neural networks (NNs), particularly multilayer perceptrons (MLPs), have emerged as more flexible alternatives due to their ability to approximate complex nonlinear behaviors efficiently \cite{mai2022robust,shojaeefard2013modelling,kabir2021failure}.

MLPs have been extensively employed in structural engineering applications for predicting structural strength and failure states, enabled by their universal approximation capabilities \cite{hornik1989multilayer,papadrakakis1998structural,bisagni2002post,sun2021prediction}. More recently, convolutional neural networks (CNNs) and generative adversarial networks (GANs) have extended the capabilities of the neural networks, effectively capturing structural details in 2D or 3D grid-based representations such as composite material characteristics \cite{ramkumar2021unconventional} and flexoelectric structural responses \cite{wang2023cnn}. However, since the above neural networks require fixed-size input vectors, they are not well suited for ship components such as stiffened panels, whose stiffener and frame layouts can vary greatly, although some initial efforts have been made for simpler geometries \cite{mokhtari2025comparison}. Since each unique geometry would require a different input vector size, these methods cannot generalize across diverse designs in an efficient optimization framework.

Graph neural networks (GNNs) provide an advanced solution for modeling non-Euclidean data structures prevalent in real-world systems, such as computer vision \cite{xu2017scene}, chemistry \cite{gilmer2017neural}, and biology \cite{fout2017protein}. GNNs have recently demonstrated significant advantages in engineering applications, for example, GNN-based reduced order models (ROMs) have effectively replaced costly computational fluid dynamics (CFD) simulations \cite{lino2022multi,pfaff2020learning,gao2022finite}. In structural mechanics, GNNs have efficiently represented truss structures, metamaterials, and lattice topologies by treating joints as nodes and structural elements as edges, effectively capturing complex structural interactions \cite{zheng2023tso,chou2024structgnn,xue2023learning,jiang2024gnns,jain2024latticegraphnet}. Cai and Jelovica \cite{cai2024efficient,cai2025heterogeneous} further demonstrated GNN-based approaches to model stiffened panels effectively using homogeneous and heterogeneous graph representations.

Recent advancements have also seen GNNs integrated into optimization and physics-informed frameworks. For structural design, GNNs have facilitated automated layout generation and optimization through differentiable frameworks and intelligent synthesis methods, significantly enhancing computational efficiency \cite{zhang2024end,zhao2023intelligent,li2023automated,zhang2024differentiable}. On the analysis side, physics-informed neural networks (PINNs) incorporate fundamental mechanical principles directly into the neural network architecture or loss functions, reducing data dependency while improving predictive fidelity \cite{raissi2019physics,song2023elastic,parisi2024use}. However, PINNs often require continuous and smooth representations of the governing partial differential equations, and their implementation can become highly complex for structures with discontinuous geometry, such as ship hull girders.

Building upon these recent advances, the present study introduces a novel hybrid framework that integrates ESL-based homogenization with a heterogeneous graph transformer (HGT) surrogate model (a subtype of GNN) for ship hull-girder stress and displacement analysis. The proposed approach enables accurate prediction of stress and displacement fields across the entire hull girder using an HGT trained exclusively on panel-level data, which is computationally more efficient than training on full 3D global models. Once trained, the HGT can predict the structural response of every panel in a hull girder. HGT is used since it demonstrated strong performance in capturing complex response of stiffened panels with non-uniform boundary kinematics \cite{cai2025heterogeneous}. Once the HGT is trained, the hull girder’s detailed response can be evaluated rapidly by performing only the computationally inexpensive global ESL analysis. This capability makes the framework particularly suitable for optimization at the initial design stage, where fast, high-fidelity hull girder analysis is essential. The framework’s effectiveness is demonstrated on three distinct hull girder structures, showing accurate stress and displacement predictions across the entire girders with a range of panel cross-section geometries.

\section{Methodology}

\subsection{Hybrid ship structural analysis framework}\label{sec2_4}

\begin{figure}[ht]
	\centering
	\includegraphics[width=0.9\linewidth]{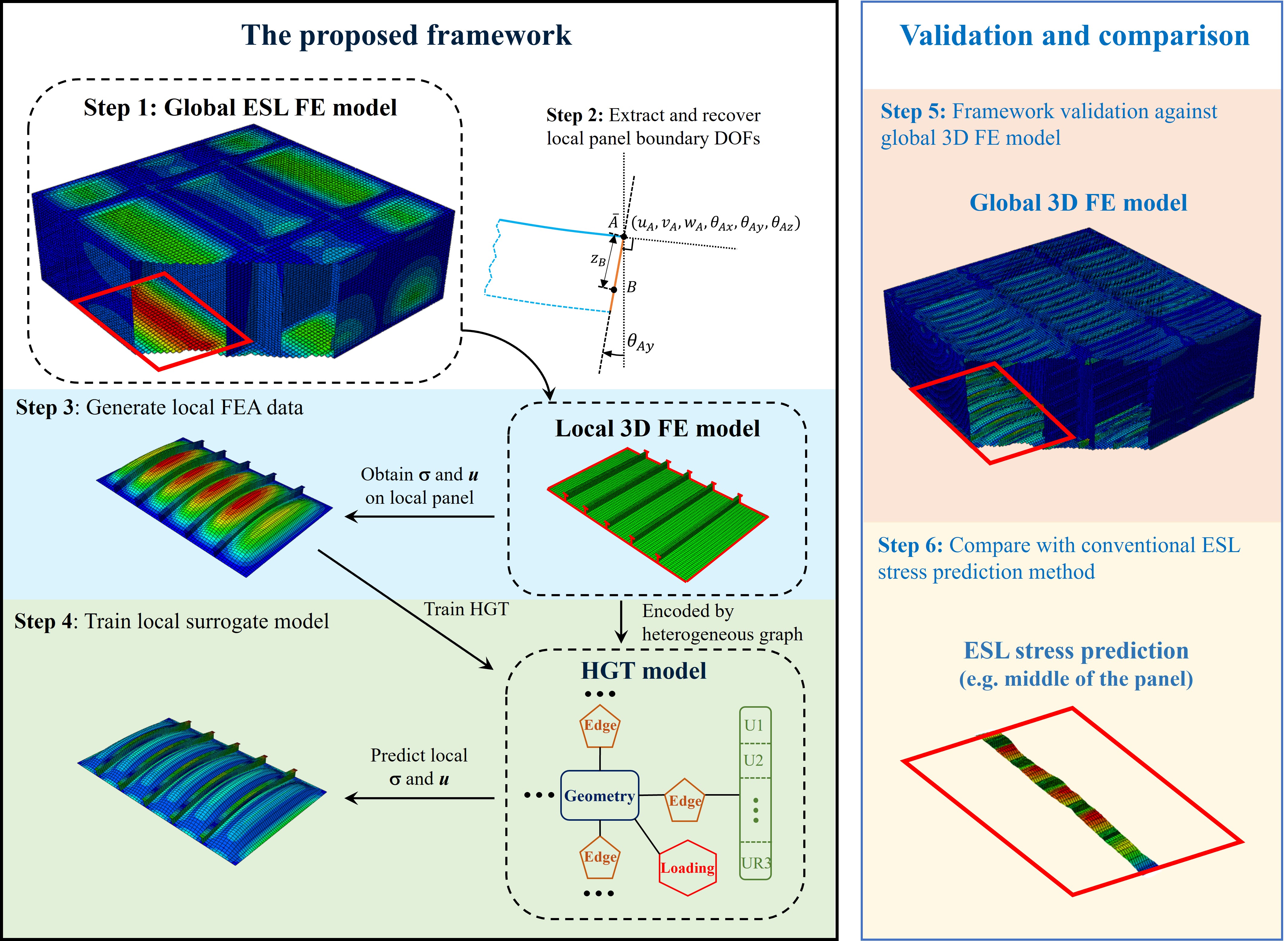}
	\caption{The proposed hybrid framework for von Mises stress and displacement prediction in hull girders, from the viewpoint of model development, training and validation. }
	\label{fig: GNN-ESL-framework}
\end{figure}

We propose a hybrid framework that achieves both computational efficiency at the global scale and high fidelity at the local panel level for ship hull analysis. This is accomplished by coupling an equivalent single-layer (ESL) coarse-mesh finite element model for global analysis with a heterogeneous graph transformer (HGT) deep learning model as a fast surrogate for local analysis. The global ESL model provides information on panel edge displacements and rotations which are used to recover detailed boundary DOFs for local analysis. The reconstructed DOFs, together with panel geometry and loading, serve as inputs to a heterogeneous graph transformer (HGT), a subtype of GNN, which rapidly and accurately predicts the detailed stress and displacement fields for any panel within the hull girder. The HGT is trained using high-fidelity 3D panel finite element model with reconstructed boundary conditions. The overall process can be separated into an offline development and training phase (Steps 1–4), and an online testing and deployment phase (Steps 5–6). The key steps are outlined below and illustrated in Fig.~\ref{fig: GNN-ESL-framework}.

\begin{description}
    \item[Step 1:] Construct coarse-mesh global ESL model: The entire hull girder (taken here as a box beam) is represented using a coarse-mesh ESL model, where stiffened panels are replaced by homogenized elements. This initial FE analysis effectively provides the global (and average) displacement field of the entire structure.

    \item[Step 2:] Extract and recover local panel boundary DOFs: The global displacement fields from the coarse-mesh ESL model are used to reconstruct the detailed DOFs along the panel's boundary for each stiffened panel, including plate, stiffener web, and flange edges. This boundary recovery procedure is detailed in Section~\ref{sec2_2}.

    \item[Step 3:] Generate high-fidelity local FEA data: The recovered boundary DOFs from Step 2 are applied as boundary conditions to a fine-mesh finite element model of a single stiffened panel. This panel-level local FEA provides high-fidelity stress and displacement fields for each stiffened panel, which are used for training the HGT surrogate model.

    \item[Step 4:] Train the local surrogate model (HGT): The local FEA results (stress and displacement fields) are paired with the corresponding panel geometries, external loadings, and recovered boundary conditions to form a heterogeneous graph dataset (Section~\ref{sec2_3_1}). The HGT surrogate model (Section~\ref{sec2_3_2}) is then trained to predict the local response based on this dataset.

    \item[Step 5 (Framework performance validation):] Stress and displacement predictions of the hybrid framework are validated against a complete, full-detail global FEA of the original structure to measure the total end-to-end error.

    \item[Step 6 (Local model comparison):] The accuracy of the HGT surrogate is compared against the conventional ESL stress prediction method (Section~\ref{sec2_1_2}) to qualitatively and quantitatively assess the performance gain of the HGT-based approach for local stress analysis.    
\end{description}

Steps 3 and 4 are omitted once the HGT is trained, or in other words, when the framework is ready for practical use. The HGT receives boundary DOFs, geometrical dimensions and loading as inputs and outputs stress or displacement fields of a stiffened panel.

\subsection{Equivalent single-layer model (ESL)}\label{sec2_1}
\subsubsection{Kinematics and constitutive equations}\label{sec2_1_1}

\begin{figure}
    \centering
    \includegraphics[width=0.6\linewidth]{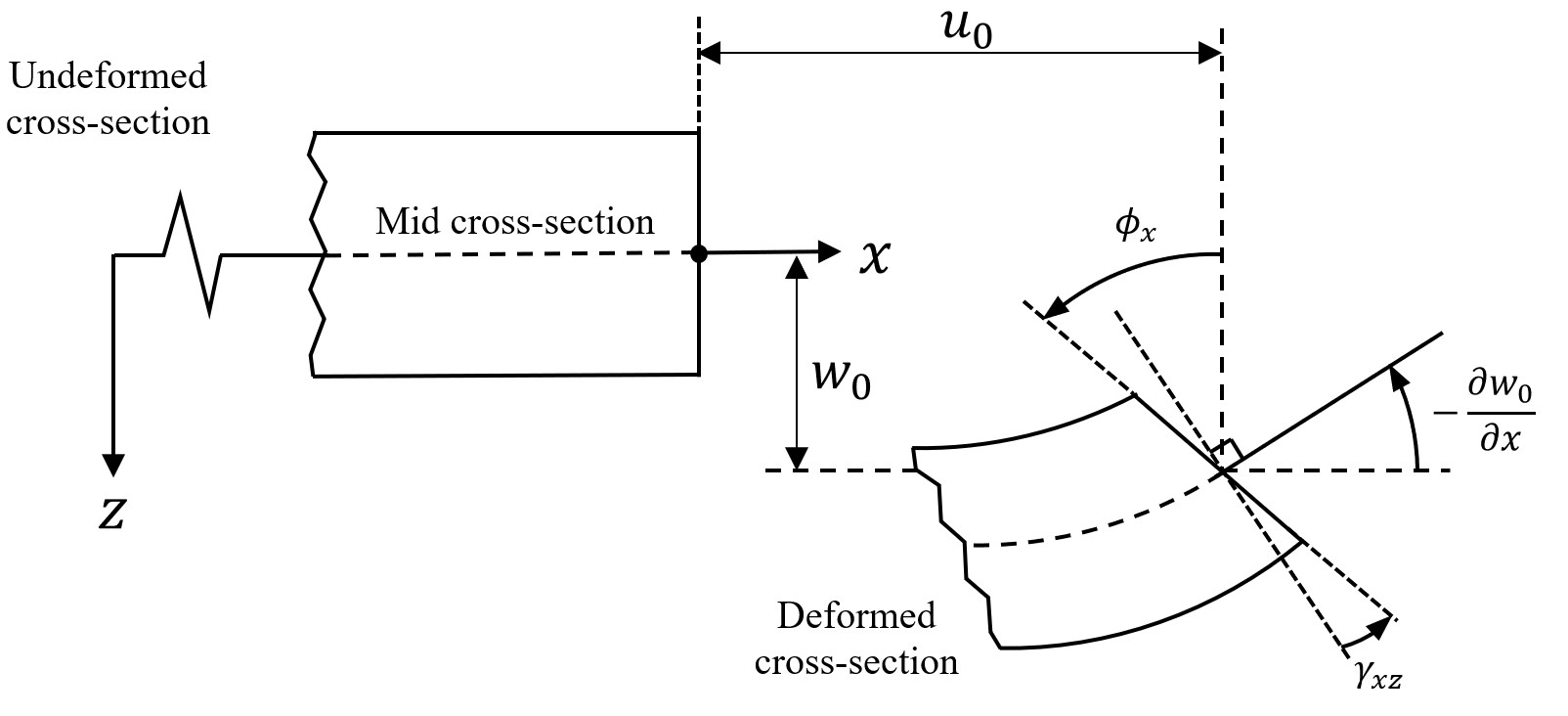}
    \caption{Undeformed and deformed geometries of a plate under the assumption of the first-order shear deformation theory (FSDT) \cite{putranto2022application}}
    \label{fig: FSDT}
\end{figure}

The ESL is used to predict the stress resultants (forces and moments) and displacement field of a hull girder using the first-order shear deformation theory (FSDT), which allows the transverse normals to deviate from being perpendicular to the mid-plane of the homogenized panel after deformation, as shown in Fig. \ref{fig: FSDT}. According to the FSDT assumptions, the displacement field of a plate can be defined as:

\begin{equation}\label{Eq: disp field FSDT}
    \begin{aligned}
        u(x, y, z, t) &= u_0(x, y, t) + z \phi_x(x, y, t), \\
        v(x, y, z, t) &= v_0(x, y, t) + z \phi_y(x, y, t), \\
        w(x, y, z, t) &= w_0(x, y, t),
    \end{aligned}
\end{equation}

\noindent where subscript 0 denotes the geometrical mid-plane displacements. The coordinate system for the stiffened panels in this study is shown in Fig.~\ref{Fig: ESL DOF reconstruct}.
Considering geometrical nonlinearity with moderate rotation but small strains, von-Karman nonlinear strain field can be expressed by:

\begin{equation}\label{Eq: strain field}
    \left\{
    \begin{array}{c}
        \varepsilon_{xx} \\
        \varepsilon_{yy} \\
        \gamma_{yz} \\
        \gamma_{xz} \\
        \gamma_{xy}
    \end{array}
    \right\}
    =
    \left\{
    \begin{array}{c}
        \varepsilon_{xx}^0 \\
        \varepsilon_{yy}^0 \\
        \gamma_{yz}^0 \\
        \gamma_{xz}^0 \\
        \gamma_{xy}^0
    \end{array}
    \right\}
    +
    z
    \left\{
    \begin{array}{c}
        \varepsilon_{xx}^1 \\
        \varepsilon_{yy}^1 \\
        \gamma_{yz}^1 \\
        \gamma_{xz}^1 \\
        \gamma_{xy}^1
    \end{array}
    \right\},
\end{equation}

\begin{equation}\label{Eq: strain field 0}
    \left\{ \varepsilon^0 \right\}
    =
    \left\{
    \begin{array}{c}
        \varepsilon_{xx}^0 \\
        \varepsilon_{yy}^0 \\
        \gamma_{yz}^0 \\
        \gamma_{xz}^0 \\
        \gamma_{xy}^0
    \end{array}
    \right\}
    =
    \left\{
    \begin{array}{c}
        \frac{\partial u_0}{\partial x} + \frac{1}{2} \left( \frac{\partial w_0}{\partial x} \right)^2 \\
        \frac{\partial v_0}{\partial y} + \frac{1}{2} \left( \frac{\partial w_0}{\partial y} \right)^2 \\
        \frac{\partial w_0}{\partial y} + \phi_y \\
        \frac{\partial w_0}{\partial x} + \phi_x \\
        \frac{\partial u_0}{\partial y} + \frac{\partial v_0}{\partial x} + \frac{\partial w_0}{\partial x} \frac{\partial w_0}{\partial y}
    \end{array}
    \right\},
\end{equation}

\begin{equation}
    \left\{ \varepsilon^1 \right\}
    =
    \left\{
    \begin{array}{c}
        \varepsilon_{xx}^1 \\
        \varepsilon_{yy}^1 \\
        \gamma_{yz}^1 \\
        \gamma_{xz}^1 \\
        \gamma_{xy}^1
    \end{array}
    \right\}
    =
    \left\{
    \begin{array}{c}
        \frac{\partial \phi_x}{\partial x} \\
        \frac{\partial \phi_y}{\partial y} \\
        0 \\
        0 \\
        \frac{\partial \phi_x}{\partial y} + \frac{\partial \phi_y}{\partial x}
    \end{array}
    \right\}.
\end{equation}\label{Eq: strain field 1}

\noindent where the strains are composed of extensional($^0$) and bending ($^1$) components. The constitutive equations are given by:

\begin{equation}\label{Eq: ABD}
    \left\{
    \begin{array}{c}
        N_{xx} \\
        N_{yy} \\
        N_{xy} \\
        M_{xx} \\
        M_{yy} \\
        M_{xy}
    \end{array}
    \right\}
    =
    \begin{bmatrix}
        A_{11} & A_{12} & 0 & B_{11} & B_{12} & 0 \\
        A_{21} & A_{22} & 0 & B_{21} & B_{22} & 0 \\
        0 & 0 & A_{33} & 0 & 0 & B_{33} \\
        C_{11} & C_{12} & 0 & D_{11} & D_{12} & 0 \\
        C_{21} & C_{22} & 0 & D_{21} & D_{22} & 0 \\
        0 & 0 & C_{33} & 0 & 0 & D_{33}
    \end{bmatrix}
    \left\{
    \begin{array}{c}
        \varepsilon_{xx}^0 \\
        \varepsilon_{yy}^0 \\
        \gamma_{xy}^0 \\
        \varepsilon_{xx}^1 \\
        \varepsilon_{yy}^1 \\
        \gamma_{xy}^1
    \end{array}
    \right\},
\end{equation}

\noindent where $N_{xx}, N_{yy}$ are the membrane forces, $N_{xy}$ is the shear force, $M_{xx}, M_{yy}$ are the bending moments, and $M_{xy}$ is the torsional moment. Matrices [A], [B], [C], and [D] are the stiffness matrices of the stiffened panel, obtained by integrating material position through the thickness. Additionally, the transverse shear strains are assumed to be constant through the thickness in FSDT. The relationship between the shear forces and average shear strains can be written as:

\begin{equation}
    \left\{
    \begin{array}{c}
        Q_x \\
        Q_y
    \end{array}
    \right\}
    =
    \begin{bmatrix}
        D_{Qx} & 0 \\
        0 & D_{Qy}
    \end{bmatrix}
    \left\{
    \begin{array}{c}
        \gamma_{xz} \\
        \gamma_{yz}
    \end{array}
    \right\},
\end{equation}

\begin{align}
D_{Qx} = k \cdot G \cdot \frac{t_w}{s} \cdot h_w \label{eq:DQx}\\
D_{Qy} = k \cdot G \cdot t_p           \label{eq:DQy}
\end{align}

\noindent where \(D_{Qx}\) and \(D_{Qy}\) represent the transverse shear stiffnesses in the stiffener direction and in the direction transverse to the stiffener, respectively. \(D_{Qy}\) is calculated following the approach proposed by Avi et al.~\cite{avi2015equivalent}, while \(D_{Qx}\) is simplified based on that work by taking only the contribution of the web, which was found sufficient in the preliminary study here, since the original formulation was developed for bulb profiles commonly used in shipbuilding, whereas the present study considers T‑profiles. \(k\) is the shear‑correction factor, taken as \(5/6\) following literature~\cite{cai2025heterogeneous}, \(G\) is the shear modulus of the material, \(t_w\) is the web thickness, \(h_w\) is the web height, and \(t_p\) is the plate thickness.

\subsubsection{Stress prediction using ESL and beam theory}\label{sec2_1_2}

Forces and moments obtained through the global ESL model can be used to approximate stress values in a panel by bringing back information about panel geometry and stiffness. However, this approach has certain limitations and is used for comparison with other approaches (the proposed framework and 3D model), not for training the HGT. 

The global ESL FE model provides the averaged stress field $\sigma_{\mathrm{av}}$ based on homogenized stiffnesses. To accurately estimate the total stress field $\sigma_{\mathrm{tot}}$ and stress variation at the plate surfaces, the local plate bending stress $\sigma_Q$ that occurs between stiffeners due to lateral pressure can be superimposed onto the global solution.

The stress field of the stiffened panel can be approximated through some assumptions. Assuming that the stiffeners do not deflect nor rotate significantly during panel bending \cite{metsala2016geometrically}, the plate segment between two adjacent stiffeners in the middle of the stiffener span can be treated as a beam undergoing bending with boundary conditions set as fixed-fixed. In each bay of width \(l\), subjected to a line loading \(q\), the bending moment per unit width varies with the local coordinate \( y \in [0, s] \) (measured from the fixed edge or a stiffener), which can be represented as:
\begin{equation}\label{Eq: M_FF_uniform}
	M_{\mathrm{uniform}}(y) \;=\; -\frac{q\,l}{2}\,y \;+\; \frac{q}{2}\,y^2 \;+\; \frac{q\,l^2}{12}\,.
\end{equation}

Additionally, for plates subjected to trapezoidal load and partially distributed triangular load, the corresponding bending moment per unit width with respect to local coordinate \( y \in [0, s] \) is  \cite{young2002roark}:

\begin{equation}\label{Eq: M_FF_trapezoidal}
M_{\text{trapezoidal}}(y)
=
\frac{q_{1}\,l^{2}}{20}
+ \,\frac{q_{2}\,l^{2}}{30}
-\,\frac{7\,q_{1} + 3\,q_{2}}{20}\;l\,y
+\,\frac{q_{1}\,y^{2}}{2}
+\,\frac{(q_{2}-q_{1})\,y^{3}}{6\,l}\,.
\end{equation}

\begin{equation}\label{Eq: M_FF_traingular}
M_{\text{triangular}}(y)
=
\begin{cases}
  \dfrac{q_{2}\,(l - a)^{3}\,\bigl(3\,l\,a - 9\,l\,y - 6\,a\,y + 2\,l^{2}\bigr)}{60\,l^{3}},
    & x \le a,\\[10pt]
  \dfrac{q_{2}}{60\,l^{3}\,(l - a)}
    \begin{multlined}[t]
      \bigl(2\,l^{6} -5\,l^{5}a -9\,l^{5}y +30\,l^{4}a\,y -30\,l^{3}a\,y^{2} \\
      \quad +\,10\,l^{3}y^{3} -10\,l^{2}a^{4} +3\,l\,a^{5} +15\,l\,a^{4}y -6\,a^{5}y\bigr)
    \end{multlined},
    & a < y \le l;
\end{cases}
\end{equation}

\noindent where $q_{1}$ and $q_{2}$ are the magnitudes of the load on the left- and right-hand limits of the distribution, respectively, and $a$ defines the starting location of the partially distributed triangular load. An illustration of the dimensions is given in Fig.~\ref{Fig: Paper3_appendix_localization_BCs}. The preliminary study presented in~\ref{app: esl_local} showed that assuming fixed–fixed boundary conditions yields the most accurate stress predictions; this assumption is therefore adopted in the rest of the article for comparison with the global 3D FE solution and the proposed framework. 

These local moments $M(y)$ are used to calculate the local stress contribution $\sigma_{Q}$ under the assumption of pure cylindrical bending. The localized normal stress contributions in \(y\)- and \(x\)-directions are:
\begin{align}
	\sigma_{Qy}(z) &= -\,z\,\frac{12\,M}{t_f^3}\,, \label{eq:Sigma_Qy}\\
	\sigma_{Qx}(z) &= \nu\,\sigma_{Qy}(z)\,. \label{eq:Sigma_Qx}
\end{align}
where M is the local bending moment (Eqs.~\ref{Eq: M_FF_uniform}--\ref{Eq: M_FF_traingular}), $\nu$ indicates the Poisson’s ratio and $t_f$ represents the plate thickness. $z$ is the thickness coordinate, positive toward the top surface.

The mid-plane strains $\varepsilon^0$ and curvatures $\varepsilon^1$ are derived by inverting the classical $\text{ABD}$ constitutive relation using the resultant membrane forces $N$ and moments $M$:
\begin{equation}\label{eq:abd-inverse}
	\begin{pmatrix}
		\varepsilon_{xx}^{(0)} \\[2pt]
		\varepsilon_{yy}^{(0)} \\[2pt]
		\gamma_{xy}^{(0)} \\[2pt]
		\varepsilon_{xx}^{(1)} \\[2pt]
		\varepsilon_{yy}^{(1)} \\[2pt]
		\gamma_{xy}^{(1)}
	\end{pmatrix}
	=
	\begin{pmatrix}
		A_{11} & A_{12} & 0 & B_{11} & B_{12} & 0 \\
		A_{21} & A_{22} & 0 & B_{21} & B_{22} & 0 \\
		0 & 0 & A_{33} & 0 & 0 & B_{33} \\
		C_{11} & C_{12} & 0 & D_{11} & D_{12} & 0 \\
		C_{21} & C_{22} & 0 & D_{21} & D_{22} & 0 \\
		0 & 0 & C_{33} & 0 & 0 & D_{33}
	\end{pmatrix}^{-1}
	\begin{pmatrix}
		N_{11}\\[2pt]
		N_{22}\\[2pt]
		N_{12}\\[2pt]
		M_{11}\\[2pt]
		M_{22}\\[2pt]
		M_{12}
	\end{pmatrix}.
\end{equation}

The normal and shear strains are calculated as:

\begin{equation}\label{Eq: ESL strain}
\begin{Bmatrix}
\varepsilon_{xx}\\
\varepsilon_{yy}\\
\gamma_{xy}
\end{Bmatrix}
=
\begin{Bmatrix}
\varepsilon_{xx}^{(0)}\\
\varepsilon_{yy}^{(0)}\\
\gamma_{xy}^{(0)}
\end{Bmatrix}
+ z 
\begin{Bmatrix}
\varepsilon_{xx}^{(1)}\\
\varepsilon_{yy}^{(1)}\\
\gamma_{xy}^{(1)}
\end{Bmatrix}
\end{equation}

By Hooke’s law, the averaged stress components are derived from the membrane forces and moments obtained from the ESL analysis:
\begin{align}
	\sigma_{xx,\mathrm{av}}(z)
	&= \frac{E}{1-\nu^2}\,\bigl(\varepsilon_{xx} + \nu\,\varepsilon_{yy}\bigr),
	\\
	\sigma_{yy,\mathrm{av}}(z)
	&= \frac{E}{1-\nu^2}\,\bigl(\varepsilon_{yy} + \nu\,\varepsilon_{xx}\bigr),
        \\
        \tau_{xy}(z)
        &= \frac{E}{2(1+\nu^)}\,\gamma_{xy}.
\end{align}

The total normal stresses are found by superimposing the global (averaged) and local contributions:
\begin{align}
	\sigma_{xx,tot}(z)
	&= \sigma_{xx,\mathrm{av}}(z) + \sigma_{Qx}(z),\\
	\sigma_{yy,tot}(z)
	&= \sigma_{yy,\mathrm{av}}(z) + \sigma_{Qy}(z).
\end{align}

Finally, the von Mises equivalent stress at any through-thickness position \(z\) is
\begin{equation}\label{eq:vonmises}
	\sigma_\mathrm{vm}(z)
	= \sqrt{\sigma_{xx,tot}^2 - \sigma_{xx,tot}\,\sigma_{yy,tot} + \sigma_{yy,tot}^2 \;+\; 3\,\tau_{xy}^2}\,.
\end{equation}

\subsection{Bridging the global-local scales through boundary reconstruction}\label{sec2_2}

The global ESL model employs homogenized shell elements to replace stiffened panels, estimating the global displacements of the hull girder structure. However, the ESL model neglects local details, necessitating a refined local analysis of stiffened panels. To bridge this gap, the HGT model introduced in Section \ref{sec2_3} is utilized, using displacements from the global ESL model as input to predict local responses (stress and displacement fields) of the stiffened panels. However, the global model provides displacement information only at the homogenized plate level, while local analysis using HGT requires detailed displacements and rotations at the stiffened panel boundaries, including plate edges, stiffener webs, and flanges.

\begin{figure}[h]%
\centering 
\includegraphics[width=0.9\textwidth]{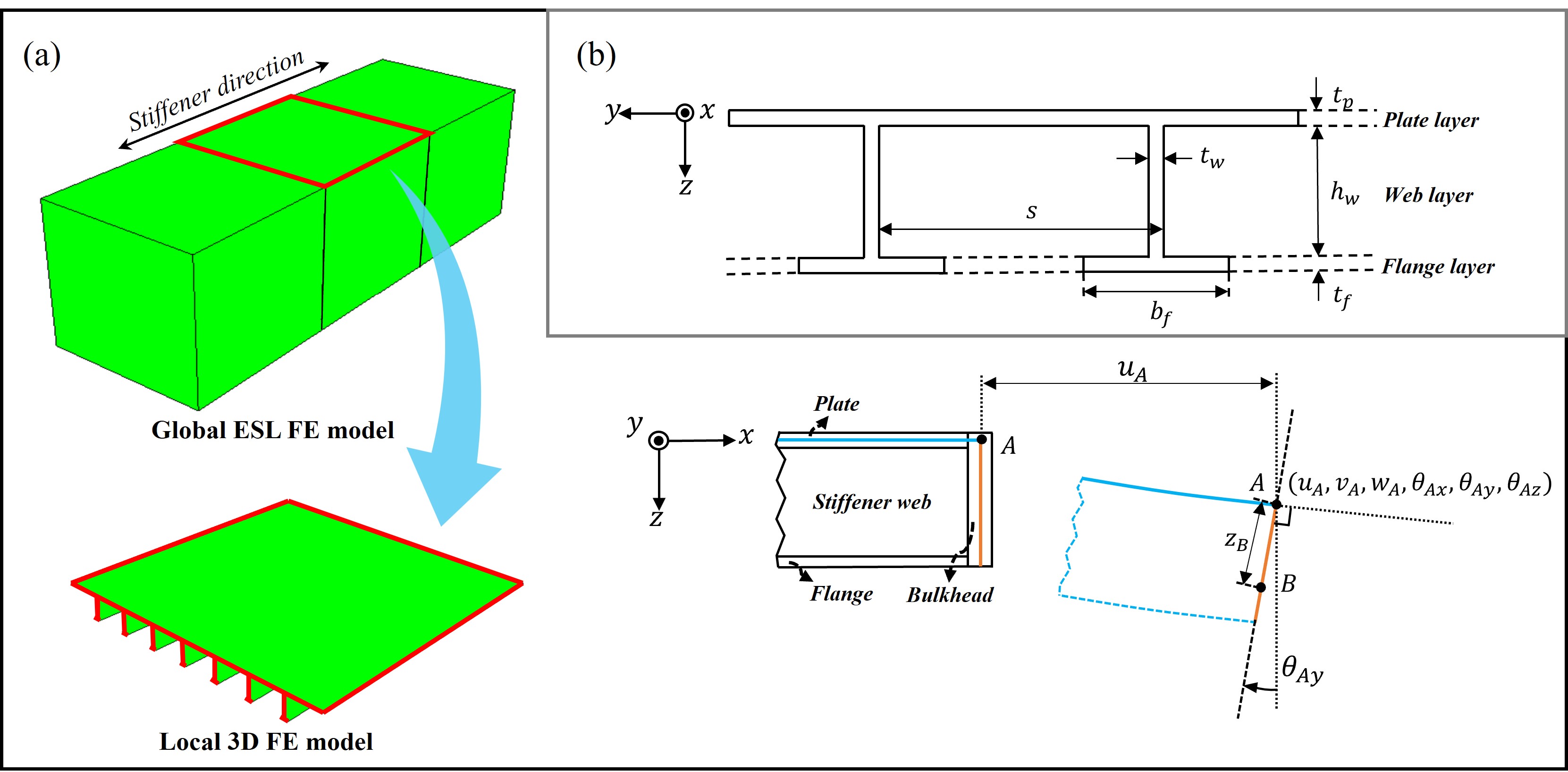} \caption{(a) Reconstructing local displacements from global ESL displacements (b) Stiffened panel cross-sectional view. }\label{Fig: ESL DOF reconstruct} 
\end{figure}

To reconstruct the boundary DOFs, it is first assumed that the displacements and rotations of the stiffener web and flange edges are equivalent to those at the corresponding positions on the bulkheads, ensuring displacement continuity. This assumption is based on the consideration that the stiffened panel is bounded by strong transverse bulkheads, which significantly influence the displacement field. However, the global ESL model does not explicitly model interactions between bulkheads and stiffeners, potentially causing incorrect rotations at bulkheads. To account for this, an adjustment procedure is proposed. This procedure is based on preliminary studies indicating that the stiffener transverse cross-section remains essentially perpendicular to its top plate at the bulkhead locations, leading to the adjustment of the translational displacements $u$ at the stiffener web and flange edges.

Fig. \ref{Fig: ESL DOF reconstruct} (a) illustrates the procedure for reconstructing the displacement \(u\). The stiffener transverse cross-section is assumed to be perpendicular to the attached plate at the bulkhead location. The solid blue and orange lines indicate the shell element locations of the homogenized plate and the adjacent bulkhead, respectively. The displacement of a point \(B\) on the web is adjusted based on the corresponding mid-plane node A. Assuming the displacements and rotations of the node A are \((u_A, v_A, w_A, \theta_{Ax}, \theta_{Ay}, \theta_{Az})\), the displacements \(u\) of the point \(B\) can be adjusted:

\begin{equation}
    \begin{aligned}
    u_B &= u_A + z_B \cdot \theta_{Ay},
    \end{aligned}
\end{equation}
where the homogenized plate surface is defined as the middle of the plate element. $z_B$ is the distance from the point $B$ to this homogenized plate surface in the $z$-direction. The transverse displacements and rotations are kept consistent with the values at the adjacent bulkhead.

\subsection{Graph-based surrogate modeling of stiffened panels}\label{sec2_3}

\subsubsection{Heterogeneous graph representation of stiffened panels}\label{sec2_3_1}

The ESL model supplies global hull displacements and rotations, which are subsequently mapped to panel boundaries and used as inputs for the HGT model. At this stage, two challenges need to be addressed to encode the local structures: (i) non-uniform boundary conditions along the edges of stiffened panels and (ii) structural variability due to different geometrical configurations. To overcome these challenges, the heterogeneous graph representation approach originally proposed in \cite{cai2025heterogeneous} is adopted, where stiffened panels are encoded as heterogeneous graphs consisting of different types of nodes and edges. That approach is further refined by combining two strategies in \cite{cai2025heterogeneous}, which allows both refined structural representation and enhanced computational efficiency. Specifically, separate node types are explicitly defined for plate edge and boundary DOFs, allowing for fine-scale differentiation. Additionally, all six boundary $\text{DOFs}$ are stored in a single node, using the approach explained below, which enhances computational efficiency by reducing the overall number of trainable parameters of the network.

\begin{figure}[ht] 
\centering 
\includegraphics[width=0.8\linewidth]{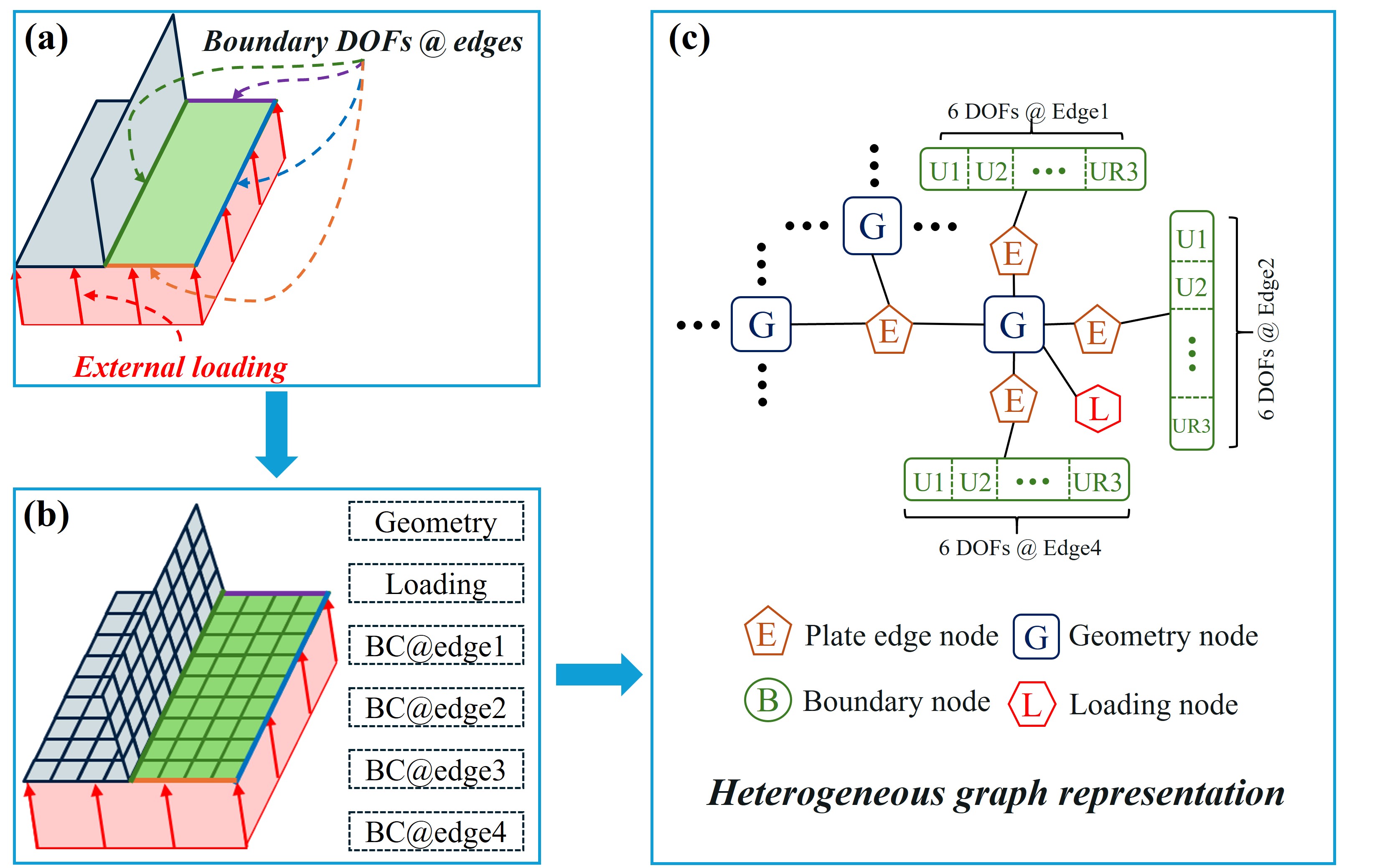} 
\caption{Heterogeneous graph representation for stiffened panels.} \label{fig: Hetero graph rep} 
\end{figure}

In this formulation, the core information of the stiffened panel, including geometric dimensions, external loadings, and boundary conditions, is encoded directly into the graph's structure and features. Specifically, each stiffened panel is subdivided into multiple rectangular plate components, which are the stiffener web, stiffener flange and plate strip between stiffeners. Separate node types are defined for plate geometry, plate edges, external loadings, and boundary conditions. Edges are constructed between these node types to capture their interactions and further differentiated based on their relative orientations (e.g., plate-to-boundary alignment). This explicit heterogeneity allows for richer structural information to be preserved. The heterogeneous graph representation used in this study is illustrated in Fig. \ref{fig: Hetero graph rep}. For the example structure shown in this figure, the corresponding heterogeneous graph consists of 3 geometry nodes, 10 plate edge nodes, 2 loading nodes, and 6 boundary nodes, assuming only bottom plate is constrained.

Geometrical parameters, which are essentially scalar values (e.g., length, thickness), are encoded as one-dimensional feature vectors. In contrast, the boundary conditions along each plate edge are spatially varying; their distributions are conceptually represented as continuous functions. To enable numerical input to the HGT, these functions are discretized by uniformly sampling 60 points along the edge. The resulting vector representation of this sampled function captures the spatial variability. Specifically, all six DOFs of the boundary condition at each of the 60 sampled points are stacked into a single feature vector, resulting in a $1 \times 360$ dimension input vector for each boundary node ($6 \text{ DOFs} \times 60 \text{ points}$). Similarly, the external loading (pressure applied to the structural unit) is represented by a $1 \times 10$ dimension input vector. This dimension captures the pressure's spatial variability transverse to the stiffener direction, based on the uniform sampling of 10 points across the width of the plate, given that in this study loading only varies in that direction.

The outputs of the HGT model are the stress and displacement fields of each rectangular plate strip. These are stored as vectors of size $1 \times 500$, which are then reshaped into $10 \times 50$ arrays to provide a two-dimensional representation of the field distribution. Specifically, 10 points are sampled transverse to the stiffener direction, and 50 points are sampled along the stiffener direction, considering the structural units are much longer in the stiffener direction than in the transverse direction.

\subsubsection{Graph-based surrogate model}\label{sec2_3_2}

To process the heterogeneous graph inputs and predict stress and displacement fields, the HGT is used~\cite{hu2020heterogeneous}, since it demonstrated the best performance across other heterogeneous graph neural networks for the current case studies during preliminary investigations. This architecture is specifically designed to handle multiple node and edge types while efficiently propagating information across heterogeneous graphs. Each HGT convolutional layer, the graph embedding of node $v$ at layer $l$, denoted $H_v^{(l)}$, is determined by the following update mechanism:
\begin{equation}
H_v^{(l)} \;=\; \sigma\bigl(\tilde{H}_v^{(l)}\bigr)\,W^A_{\mathcal{T}_V(v)} \;+\; H_v^{(l-1)},
\end{equation}
where $W^A_{\mathcal{T}_V(v)}$ represents the type‑specific weight. $\tilde{H}_v^{(l)}$ is responsible for aggregating messages received from node $v$'s neighbors:
\begin{equation}
\tilde{H}_v^{(l)} 
= 
\bigoplus_{u \in N(v)} \!\Bigl(\alpha_{u,e,v}\,\mathbf{m}_{u,e,v}^{(l)}\Bigr),
\end{equation}
with $\mathbf{m}_{u,e,v}^{(l)}$ being the multi‑head message defined as:
\begin{equation}
\mathbf{m}_{u,e,v}^{(l)} 
= 
\;\big\Vert_{i=1}^{h}
\Bigl(H_u^{(l-1)}\,W^{\mathrm{MSG}}_{\mathcal{T}_V(u)}\,W^{\mathrm{MSG}}_{\mathcal{T}_E(e)}\Bigr).
\end{equation}

Here, the parameter $h$ specifies the number of attention heads. The learned matrices $W^{\mathrm{MSG}}_{\mathcal{T}_V(u)}$ and $W^{\mathrm{MSG}}_{\mathcal{T}_E(e)}$ are distinct and specific to the source-node type and edge type, respectively. It should be noted that $u$ and $v$ are used throughout the HGT formulation to represent the source and target nodes, respectively, following the convention established in the GNN literature. These notations are distinct from the displacement components used in the mechanics model. The attention weight $\alpha_{u,e,v}$ is determined through a multi-head self-attention mechanism applied over each meta-relation tuple $(u,e,v)$:

\begin{equation}
\alpha_{u,e,v}
=
\operatorname{Softmax}\!\Bigl(\big\Vert_{i=1}^{h}\,\mathrm{ATT\text{-}Head}_i^{(l)}(u,e,v)\Bigr).
\end{equation}

Each attention head is defined by the following expression:
\begin{equation}\label{eq:att-head}
\mathrm{ATT\text{-}Head}_i^{(l)}(u,e,v)
=
\frac{\bigl(K_u^{(l)}\,W^{\mathrm{ATT}}_{\mathcal{T}_E(e)}\,Q_v^{(l)}\bigr)\,\mu_{\langle\mathcal{T}_V(u),\mathcal{T}_E(e),\mathcal{T}_V(v)\rangle}}
{\sqrt{d}}.
\end{equation}

\begin{equation*}
K_u^{(l)} = H_u^{(l)}\,W^K_{\mathcal{T}_V(u)}, 
\quad
Q_v^{(l)} = H_v^{(l)}\,W^Q_{\mathcal{T}_V(v)}.
\end{equation*}

The query $Q_v^{(l)}$ and key $K_u^{(l)}$  are first computed by projecting the hidden states of the target node $v$ and the source node $u$ using the type-specific matrices $W^{Q}_{\mathcal{T}_V(v)}$ and $W^{K}_{\mathcal{T}_V(u)}$, respectively. Additionally, HGT injects an edge‑type‑specific projection \(W^{\mathrm{ATT}}_{\mathcal{T}_E(e)}\) and a learned prior tensor \(\mu\) for each meta‑relation to better capture distributional differences.

\begin{figure}
    \centering
    \includegraphics[width=1\linewidth]{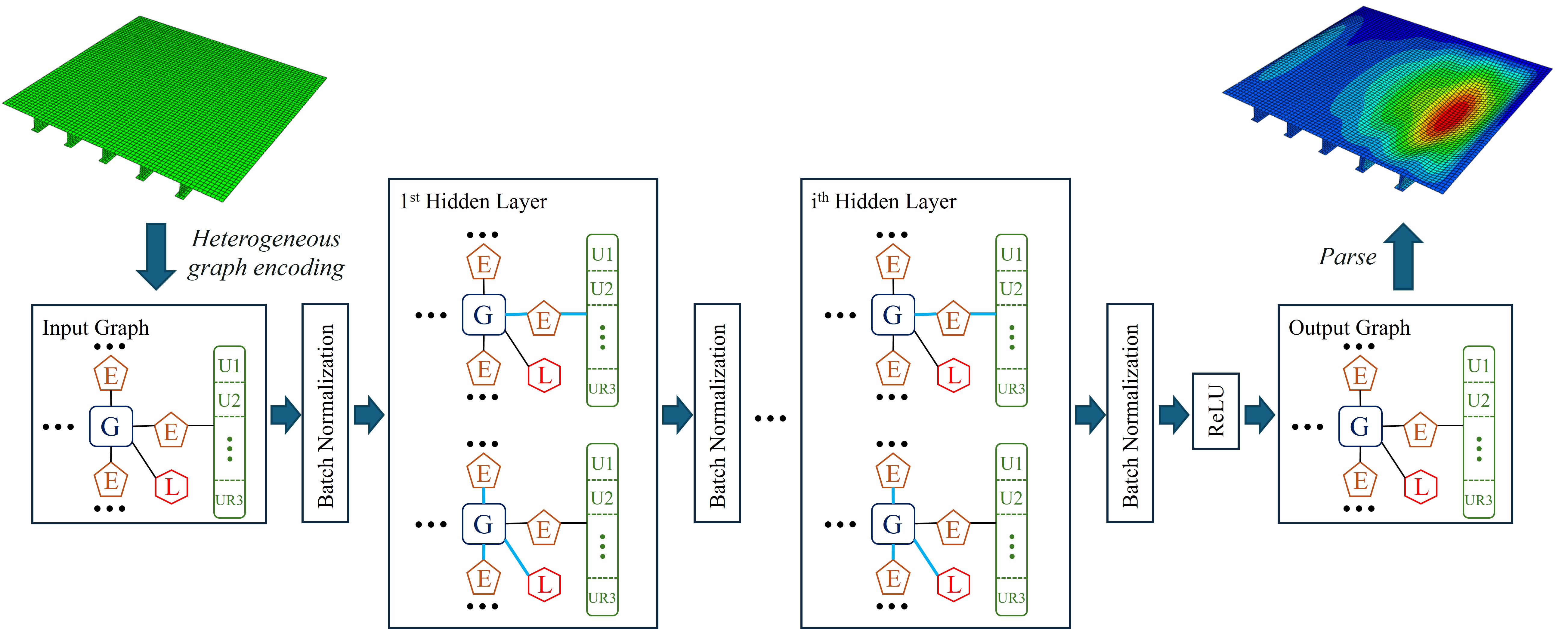}
    \caption{The architecture of the HGT model employed in this study.}
    \label{fig: HGT_architecture}
\end{figure}

Fig.~\ref{fig: HGT_architecture} represents the HGT architecture employed in this research. At the first layer, the feature space for each type of node will be projected to the same dimensional space, followed by passing through HGT layers, with batch normalization positioned after each layer to stabilize NN training. As the figure indicates, each type of node is updated by the connected nodes and edges, at each hidden layer. The hyperparameters of this model have been fine-tuned via quasi-random search. The root mean squared error (RMSE) is used as the loss function, which is calculated over all discrete prediction points across all panels in the training batch. The RMSE is defined as:
\begin{equation}
\mathrm{RMSE}
=
\sqrt{\frac{1}{n}\sum_{i=1}^n\|\mathbf{y}_i - \hat{\mathbf{y}}_i\|^2},
\label{Eq:RMSE}
\end{equation}
\noindent where $\mathbf{y}_i$ and $\hat{\mathbf{y}}_i$ are the true and predicted values (stress or displacement magnitude), respectively, at the $i$th location. The index $i$ runs over $n$, the total number of discrete prediction points across all predicted fields in a training batch. Since the output field for each rectangular plate is a $10 \times 50$ grid (500 points), the total number of sample points $n$ is calculated as the sum of $500$ for every rectangular plate across all stiffened panels in the current batch.

\section{Case studies and data preparation}\label{sec3}

The hull girder is modeled as a simplified thin-walled box girder. Three example cases with different cross-sections and loading conditions are analyzed to test the proposed framework. The first case study is a single-cell box beam subjected to uniform pressure on the top. The second case study considers a two-cell box beam structure featuring a double bottom, under four-point bending. The third and most complex scenario is a box beam with a three-cell configuration featuring two double sides, subjected to uniform pressures on the top and bottom panels, alongside partially distributed triangular pressure applied to its side panels, simulating hydrostatic loading conditions. 

\begin{figure}[ht]
	\centering
	\includegraphics[width=0.7\linewidth]{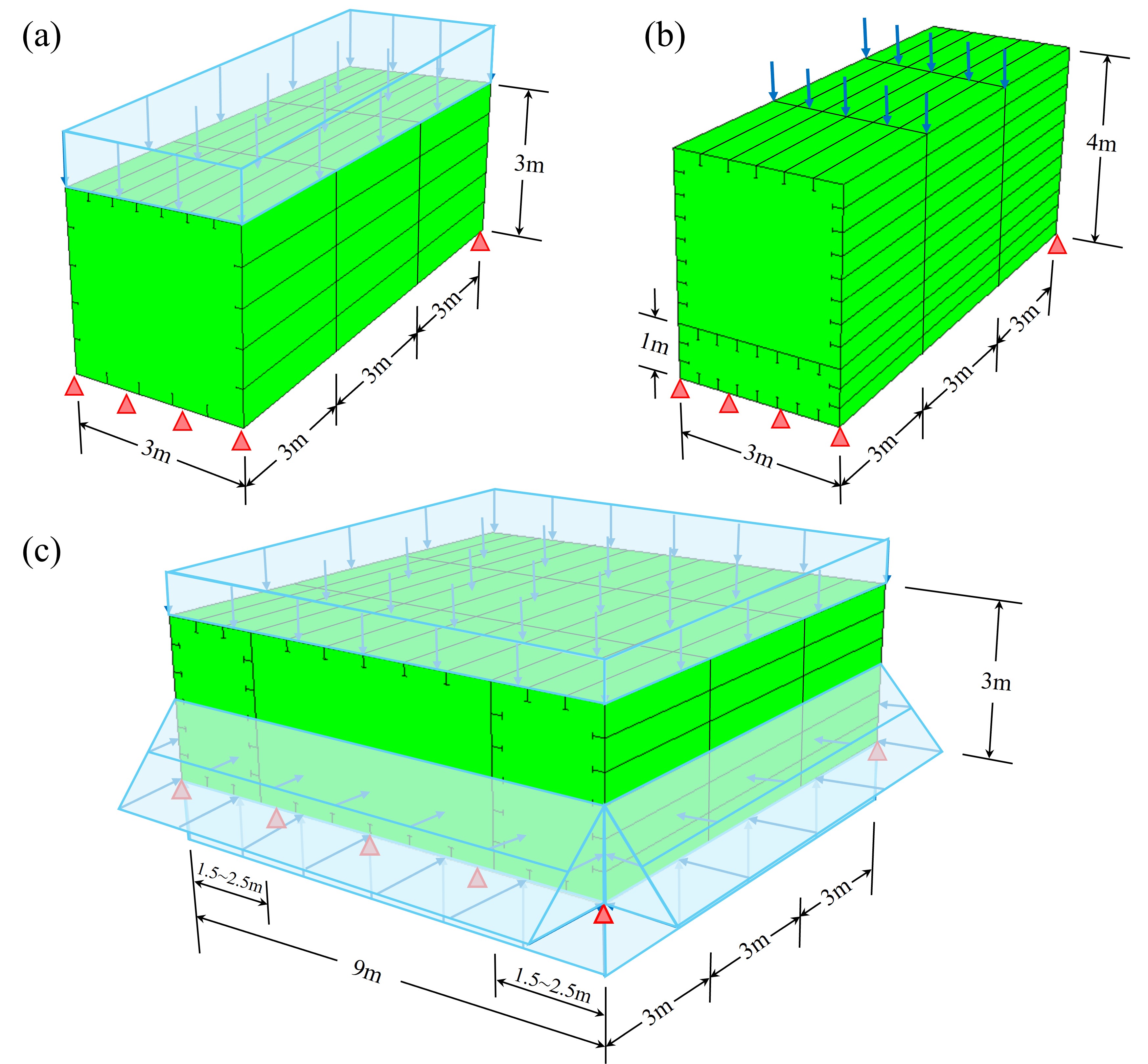}
	\caption{Geometry, loading and boundary conditions for box beams in (a) Case study 1. (b) Case study 2. (c) Case study 3.}
	\label{fig: Paper3_boxbeam_configs}
\end{figure}

The geometry of the box beams, external loading, and boundary conditions are illustrated in Fig.~\ref{fig: Paper3_boxbeam_configs}. Note that to ensure a more diverse test environment, the width of the double-sided structures in case study 3 was set to vary between $1.5\text{ m}$ and $2.5\text{ m}$. Each structure contains four equally spaced transverse bulkheads, which are modeled as 60 mm thick isotropic platesk. The primary structure, aside from the bulkheads, is composed of stiffened panels, with the top, bottom, and side panels each featuring potentially independent stiffener and plate geometries. The ranges for those geometric parameters are summarized in Table \ref{tab: var limit}.

\begin{table}[h]
\centering
\small
\caption{Lower and upper limits of geometric variables for stiffened panels in different case studies.}
\label{tab: var limit}
\begin{tabular}{@{}lccc@{}}
\toprule
Category (unit)                 & Case study 1 \& 2 & Case study 3\\ 
\midrule
Plate thickness (mm)          & 10–20        & 5–10\\ 
Web thickness (mm)      & 5–20         & 4–8\\ 
Web height (mm)         & 100–200      & 100–200\\ 
Flange thickness (mm)        & 5–20         & 4–8\\ 
Flange width (mm)             & 50–100       & 50–100\\ 
Number of stiffeners    & 2–7          & 2–7\\ 
\bottomrule
\end{tabular}
\end{table}

The applied external loadings also vary among the case studies. In the first case study, uniform pressure applied to the top of the box beam ranges from $1.11 \times 10^5$ to $3.33 \times 10^5$ Pa. The second case study introduces four-point bending, with line loads applied at the intersections of bulkheads and top panels, varying from 500 to 1500 kN/m. The third case study represents a more realistic condition with the structure immersed in water, where structural buoyancy effect has been considered. In this case study, uniform pressure loading on the top panel ranges from $6.17 \times 10^4$ to $1.85 \times 10^5$ Pa, while the water-induced pressures acting on the bottom and side panels are computed from the corresponding top panel pressures and structural weight. The box beams are assumed to be made of steel material with a density of 7850 tonnes/m\textsuperscript{3}. The material has a Young's modulus of 200 GPa and a Poisson's ratio of 0.3. 

All datasets employed for training and validating the neural networks are generated using the ABAQUS finite element software with parametric modeling. The data sensitivity analysis for the HGT model presented in~\ref{app: training_size} showed that high accuracy is achieved with 6000 data samples (stiffened panels), although already good accuracy can be achieved with only a few hundred samples. Nonetheless, the HGT used in the continuation has been trained on 6000 samples to yield high accuracy. For each case study, the 6000 data samples are partitioned into 80\% training, 10\% validation, and 10\% test sets. Since each box beam geometry is composed of multiple stiffened panels, and the number of panels varies between geometries, each box beam contributes a different quantity of training data. In this study, the $6000$ data per case study were sourced from $500$ distinct box beam geometries for case study 1, $286$ for case study 2, and $200$ for case study 3. All structures are discretized using the S4R shell element. Model accuracy was verified through mesh convergence studies, resulting in the adoption of 10 elements between stiffeners, 6 elements along the stiffener height, and 4 elements across flange widths, as validated in prior studies \cite{cai2024efficient}. The neural network training procedures are executed using the PyTorch Geometric library on a computer with an NVIDIA GeForce RTX 3090 GPU. With a maximum epoch of 500 and a batch size of 64, the training time per neural network is about 6 hours.

\section{Results}

\subsection{Error analysis}\label{sec4_1}

This section evaluates the accuracy of the proposed hybrid framework across its main stages, using the three box beam case studies introduced in Section~\ref{sec3}. For each case study, four distinct neural networks were trained separately to predict the three displacements ($u_1$, $u_2$, $u_3$) and the von Mises stress. The objective of this analysis is to identify the main sources of error, calculated as $\text{RMSE}$. Each error represents the discrepancy between a stage of the framework and a corresponding 3D FE model:
\begin{itemize}
    \item $\text{ESL}$ error (Local $\text{3D}$ $\text{FE}$ model vs. Global 3D FE model): Error arising from the ESL modeling assumptions and boundary reconstruction for the local 3D FE model. 
    \item $\text{HGT}$ error ($\text{HGT}$ vs. Local $\text{3D}$ $\text{FE}$ model): Error due to HGT modeling and training on the local 3D FE data.
    \item Final (framework) error ($\text{HGT}$ vs. Global 3D FE model): The end-to-end discrepancy of the proposed hybrid framework.
\end{itemize}

Table~\ref{tab: ESL_GNN_framework_error} summarizes these errors across all three case studies. $u_1$, $u_2$, and $u_3$ represent the displacements along the coordinate axes shown in Fig.~\ref{Fig: ESL DOF reconstruct}. Note that the $\text{ESL}$ error and $\text{HGT}$ error are not expected to sum linearly to the framework error because the $\text{RMSE}$ is a non-linear metric, and each error term represents the overall average deviation from its specific reference set.

\begin{table}[htbp]
  \centering
  \small
  \caption{RMSE of stress (MPa) and displacement (mm) at different stages of the framework for all case studies.}
  \label{tab: ESL_GNN_framework_error}
  \begin{tabular}{@{}lccc@{}}
    \toprule
    \textbf{Case} & \textbf{ESL error} & \textbf{HGT error} & \textbf{Final (framework) error} \\
    \midrule
    \textbf{Case study 1} & & & \\
    \quad von Mises stress (MPa)& 18.47 & 1.815 & 18.47 \\
    \quad $u_1$ (mm)    & 0.142 & 0.007 & 0.142 \\
    \quad $u_2$ (mm)    & 0.351 & 0.016 & 0.354 \\
    \quad $u_3$ (mm)    & 0.047 & 0.005 & 0.048 \\
    \quad Tot.\ Disp. (mm)& 0.372 & 0.013 & 0.374 \\
    \midrule
    \textbf{Case study 2} & & & \\
    \quad von Mises stress (MPa)& 2.832 & 0.262 & 2.823 \\
    \quad $u_1$ (mm)    & 0.0053 & 0.0016 & 0.0055 \\
    \quad $u_2$ (mm)    & 0.0093 & 0.0059 & 0.0102 \\
    \quad $u_3$ (mm)    & 0.0017 & 0.0013 & 0.0021 \\
    \quad Tot.\ Disp. (mm)& 0.0064 & 0.0065 & 0.0082 \\
    \midrule
    \textbf{Case study 3} & & & \\
    \quad von Mises stress (MPa)& 4.433 & 1.260 & 4.541 \\
    \quad $u_1$ (mm)    & 0.0101 & 0.0026 & 0.0100 \\
    \quad $u_2$ (mm)    & 0.0497 & 0.0108 & 0.0525 \\
    \quad $u_3$ (mm)    & 0.0258 & 0.0035 & 0.0261 \\
    \quad Tot.\ Disp. (mm)& 0.0485 & 0.0098 & 0.0504 \\
    \bottomrule
  \end{tabular}
\end{table}

As demonstrated in the table, the ESL error consistently and significantly exceeds the HGT error. For the stress field, the ESL error is an order of magnitude greater than the HGT error in case studies 1 and 2, and three times higher in case study 3. To contextualize the scale of these discrepancies, the average von Mises stresses are $59.2\text{ MPa}$, $12.66\text{ MPa}$, and $16.13\text{ MPa}$ for case studies 1, 2, and 3, respectively. The HGT error is small relative to these values, indicating a high level of accuracy by the surrogate model when predicting stress fields.

Regarding displacements, the ESL error also dominates the final framework error (contributing $>\!95\%$ of the total error) in case studies 1 and 3. Conversely, in case study 2, the $\text{ESL}$ and HGT displacement errors exhibit similar magnitudes. This result is anticipated because the four-point bending load in case study 2 primarily induces smooth, macro-scale global bending and membrane forces. The ESL method is fundamentally designed to capture displacement fields accurately under these specific loading conditions. In contrast, the pressure-dominated case studies 1 and 3 introduce significant local plate bending, resulting in larger ESL errors. Among the displacement components, $u_2$ consistently demonstrates the largest deviation across all cases, while the performance for $u_1$ and $u_3$ is comparatively better.

Overall, the data from all case studies strongly indicates that the ESL error is the primary source of total framework error, consistently dominating the end-to-end discrepancy for both stress and displacement predictions. The HGT, on the other hand, demonstrates a high level of predictive accuracy.

\subsection{Validation of the hybrid framework for the box beam segments}\label{sec4_2}

\begin{figure}[ht]
    \centering
    \includegraphics[width=0.8\linewidth]{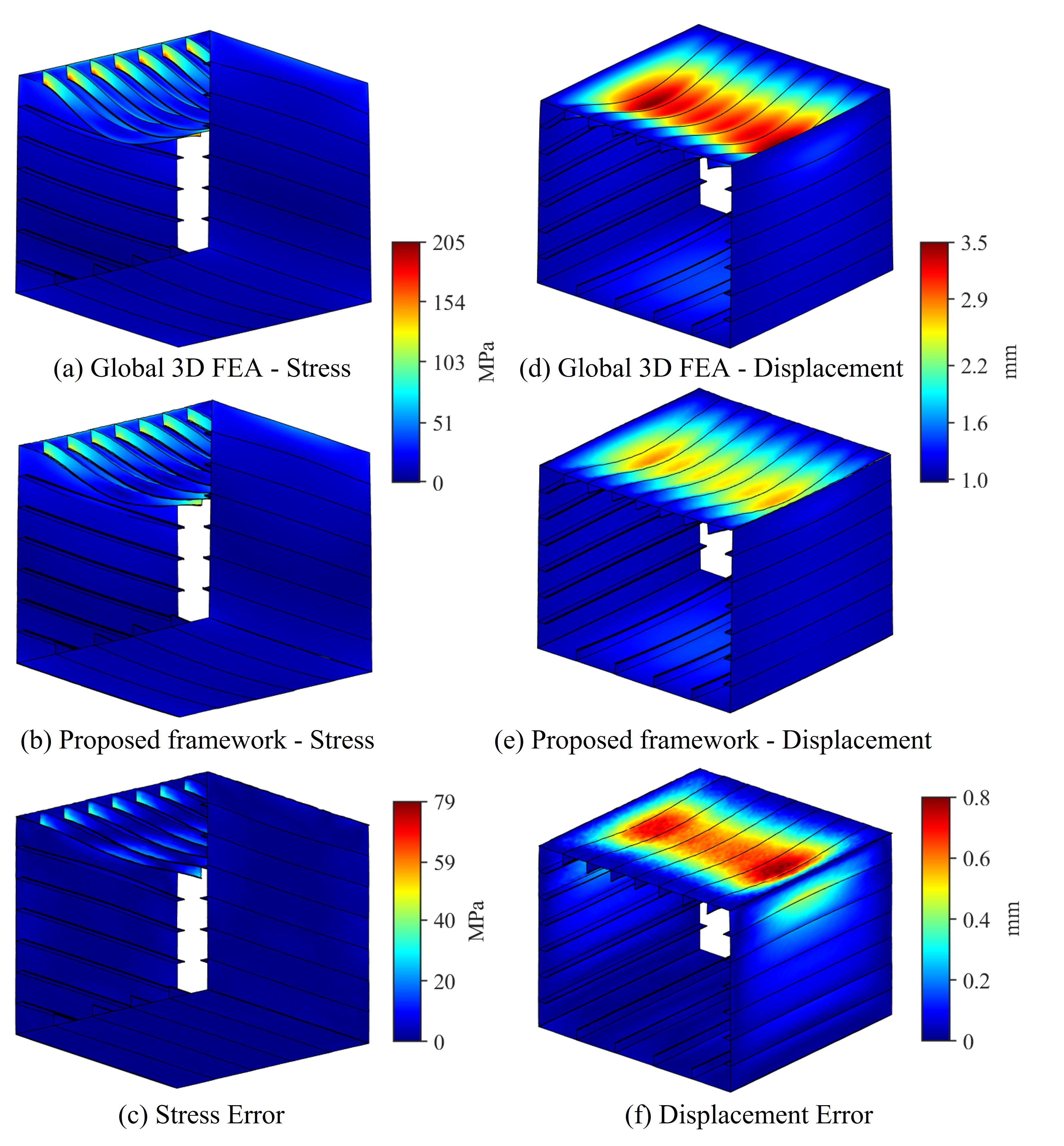}
    \caption{Contours of von Mises stress and displacement for the median-RMSE box beam segment in case study 1, showing the global 3D FE reference model, the predictions of the proposed framework, and the associated error maps.}
    \label{fig: Paper3_case1_3D_comparison}
\end{figure}

\begin{figure}[ht]
    \centering
    \includegraphics[width=0.8\linewidth]{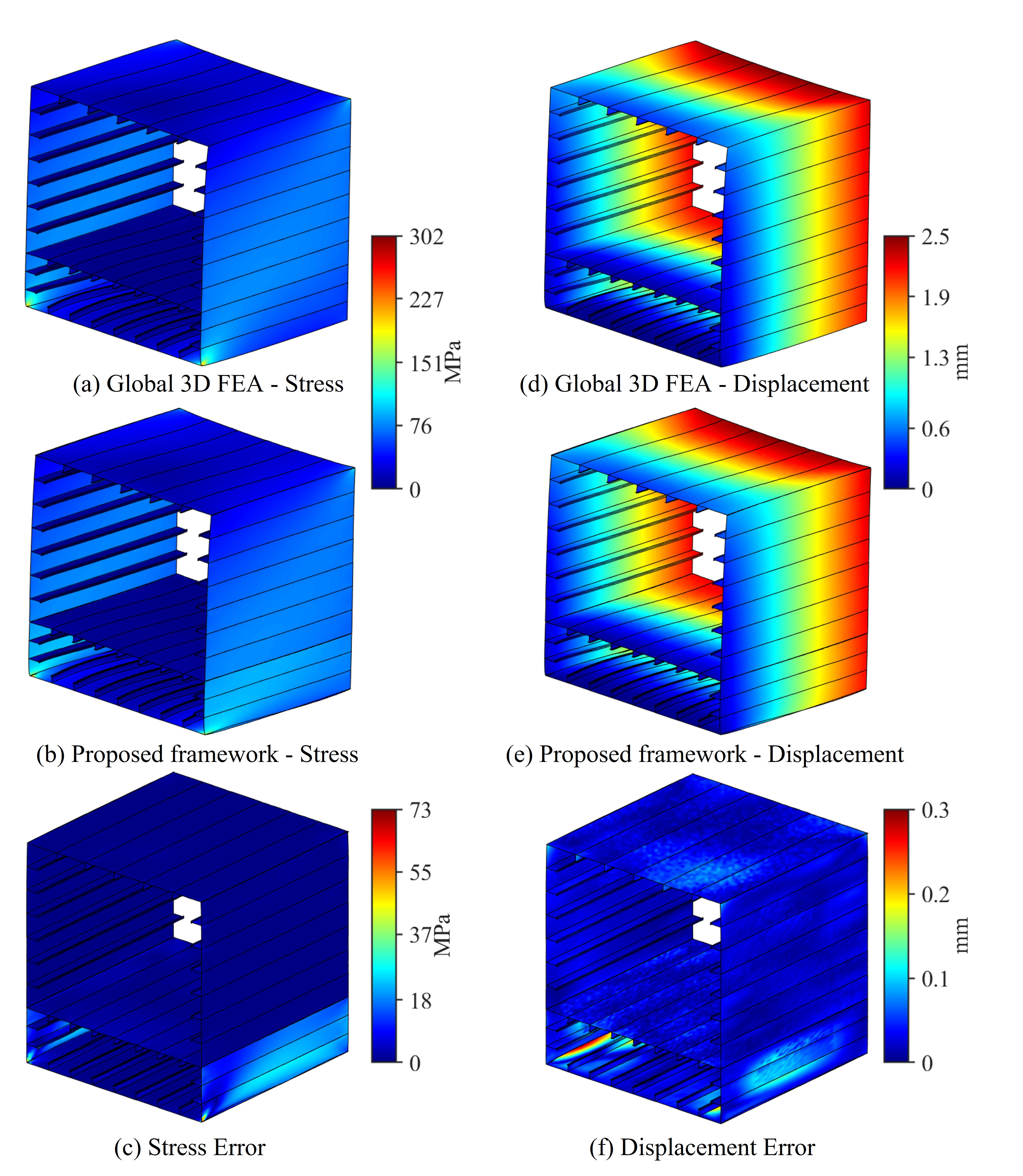}
    \caption{Contours of von Mises stress and displacement for the median-RMSE box beam segment in case study 2, showing the global 3D FE reference model, the predictions of the proposed framework, and the associated error maps.}
    \label{fig: Paper3_case2_3D_comparison}
\end{figure}

\begin{figure}[ht]
    \centering
    \includegraphics[width=0.8\linewidth]{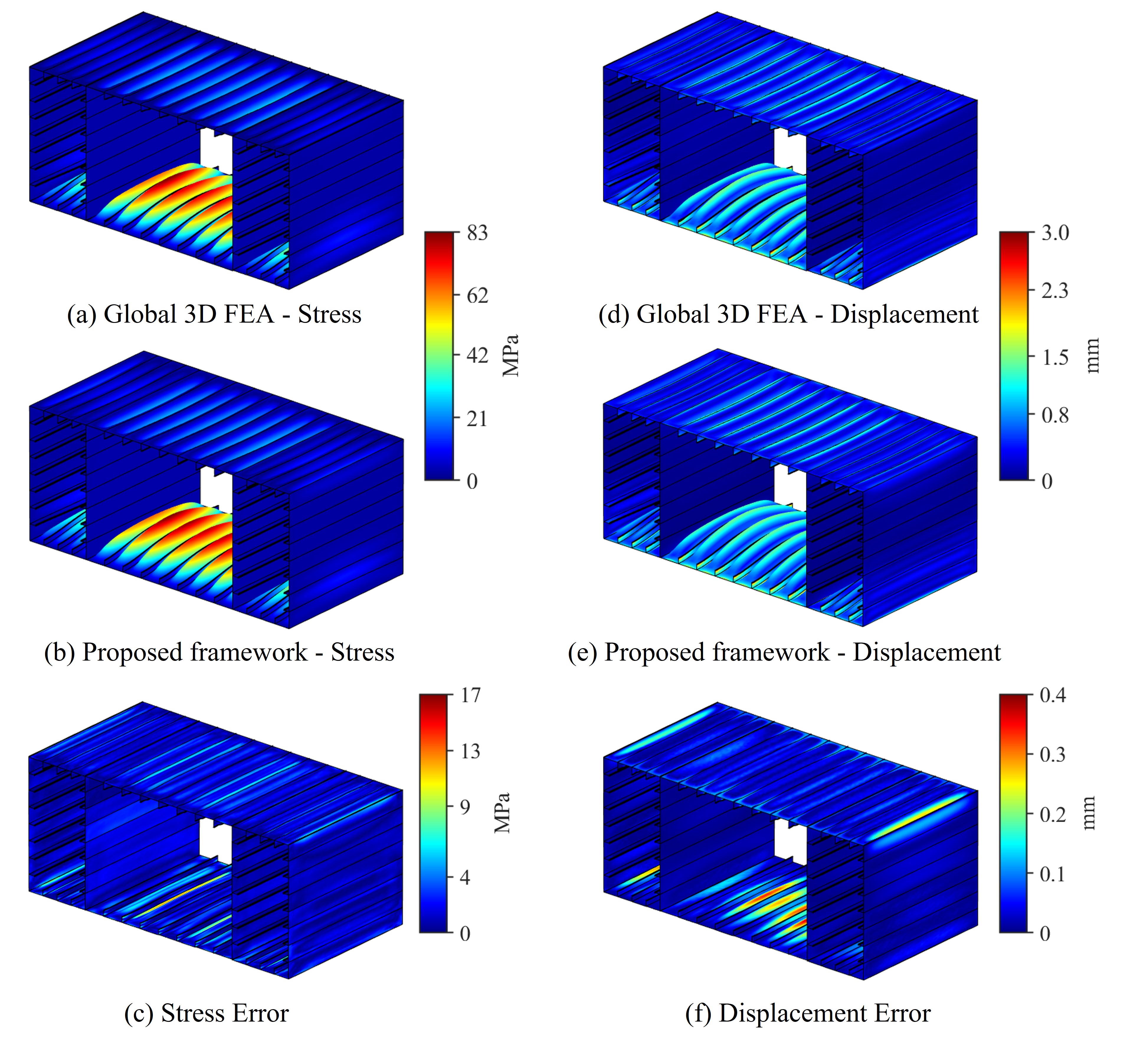}
    \caption{Contours of von Mises stress and displacement for the median-RMSE box beam segment in case study 3, showing the global 3D FE reference model, the predictions of the proposed framework, and the associated error maps.}
    \label{fig: Paper3_case3_3D_comparison}
\end{figure}

Building on the stage-wise error analysis presented in Section~\ref{sec4_1}, this section assesses the performance of the proposed hybrid framework against the global 3D FE model across a complete structural segment (or bay) of the box beam. To select a representative result, we first calculated the average $\text{RMSE}$ (for both von Mises stress and displacement) for every box beam segment in the test dataset. The segment chosen for detailed comparison is the one whose overall $\text{RMSE}$ value was closest to the statistical median of the entire test dataset. Figs.~\ref{fig: Paper3_case1_3D_comparison} through \ref{fig: Paper3_case3_3D_comparison} compare the stress and displacement fields obtained by the global 3D FE model (reference) and the proposed hybrid framework.

In case study~1 (Fig.~\ref{fig: Paper3_case1_3D_comparison}), the proposed framework predicts the stress distribution reasonably well, except for the stiffener web and flange edges, where the maximum error is around 35\%. The displacement field exhibits a noticeable underprediction at the center of the top panel, resulting in an approximate 20\% deviation. This overall error is primarily attributed to the combined effect of idealized assumptions in the $\text{ESL}$ homogenization and the boundary recovery approach. This observation aligns with the error decomposition in Section~\ref{sec4_1}, which identifies the primary discrepancies arising from the ESL and the stiffened panel boundary recovery, rather than the $\text{HGT}$ surrogate model (a finding further detailed in Section~\ref{sec4_3}).

For the represented structural segment in case study 2 (Fig.~\ref{fig: Paper3_case2_3D_comparison}), the proposed framework exhibits the smallest overall discrepancies. The deviation between the proposed framework and the global 3D $\text{FE}$ model is less than $20\text{ MPa}$ for the majority of panels within this segment. However, the maximum stress deviation reaches $73\text{ MPa}$ on the side panel of the double bottom structure. This error peak is localized at the simply supported boundary of the box beam, and the framework underestimates this stress. In addition to this error peak, the panel also exhibits moderate deviations observed near its center. This local inaccuracy can be attributed to the low number of stiffeners, as this specific panel contains only two. In contrast, the structural response of panels with more stiffeners (e.g., the top panel) is better predicted by the proposed framework. This can again be primarily attributed to the ESL model, which typically exhibits reduced accuracy when the number of stiffeners is low. Outside these specific regions, both stress and displacement predictions closely match the reference fields, with the overall error for this segment remaining below 10\%.

On the other hand, the proposed framework for the example box beam segment in case study 3 (Fig.~\ref{fig: Paper3_case3_3D_comparison}) shows a more unified performance across all panels, capturing both the stress and displacement patterns with high fidelity. The largest error for both stress and displacement predictions appears at the plate-web intersections on the bottom panel, where the applied pressure is at maximum. Nonetheless, the stress prediction remains accurate over most panels, achieving an overall $\text{RMSE}$ of $1.31\text{ MPa}$ for this entire box beam segment. The total displacement field is also predicted accurately, showing an overall $\text{RMSE}$ of $0.05\text{ mm}$. Across all presented box beam segments, the error distribution for the stress and displacement fields is different. The overall relative error for the displacement field prediction is consistently smaller than that for the stress prediction. This difference is expected, as the stress field is more complex than the displacement field, thus more challenging to train.

\subsection{Performance evaluation of the local analysis models}\label{sec4_3}

The error decomposition in Section~\ref{sec4_1} demonstrated that the $\text{HGT}$ surrogate model error has only a minor impact on the total framework discrepancy. Building on this finding, this section presents a deeper analysis of the local analysis model's performance. We first assess the stand-alone accuracy of the $\text{HGT}$ surrogate by comparing its predictions directly against its training reference, the local 3D FE model solutions (Section~\ref{sec4_3_1}). Subsequently, we compare the accuracy of this $\text{HGT}$-based approach against the conventional ESL stress prediction method (Section~\ref{sec4_3_2}), using the global 3D FE model as the reference.

\subsubsection{Prediction accuracy of the HGT surrogate model}\label{sec4_3_1}

\begin{figure}[ht]
    \centering
    \includegraphics[width=1\linewidth]{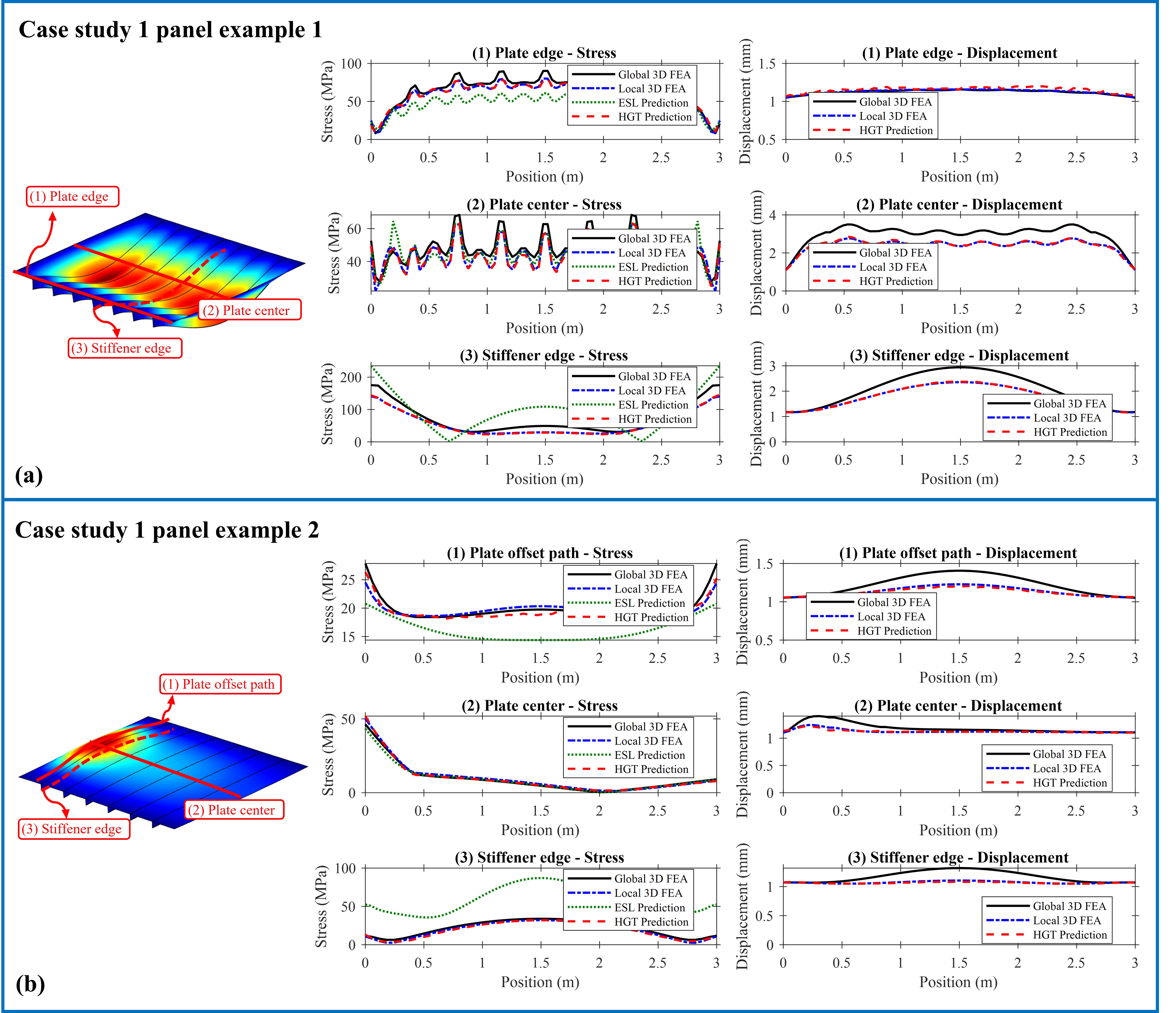}
    \caption{Comparison of the stress and total displacement for two example panels in case study 1: (a) top panel and (b) side panel. Results are shown along three prescribed paths.}
    \label{fig: Paper3_case1_panel_comparison}
\end{figure}

\begin{figure}[ht]
    \centering
    \includegraphics[width=1\linewidth]{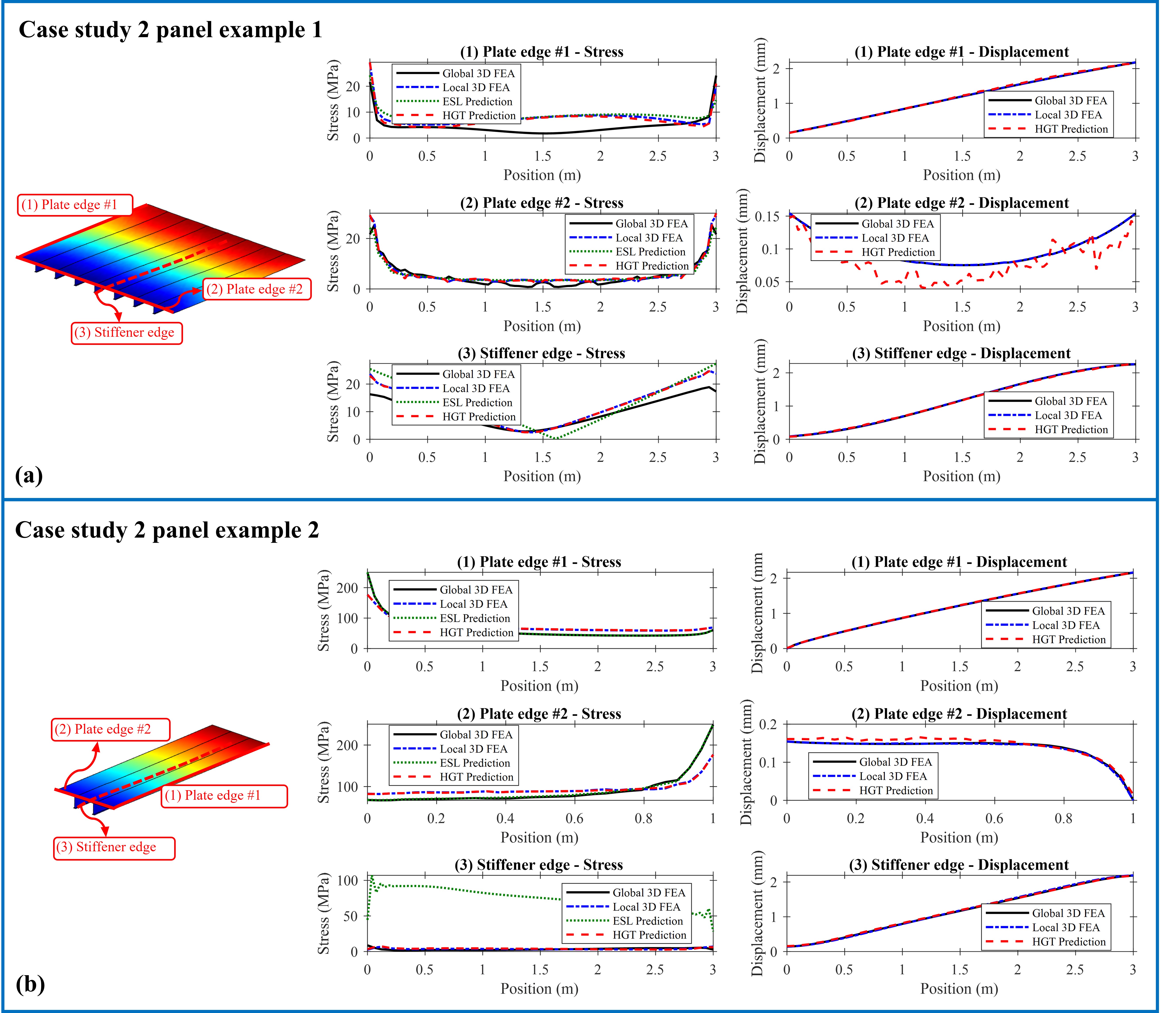}
    \caption{Comparison of the stress and total displacement for two example panels in case study 1: (a) inner-bottom panel and (b) side panel at double bottom. Results are shown along three prescribed paths.}
    \label{fig: Paper3_case2_panel_comparison}
\end{figure}

\begin{figure}[ht]
    \centering
    \includegraphics[width=1\linewidth]{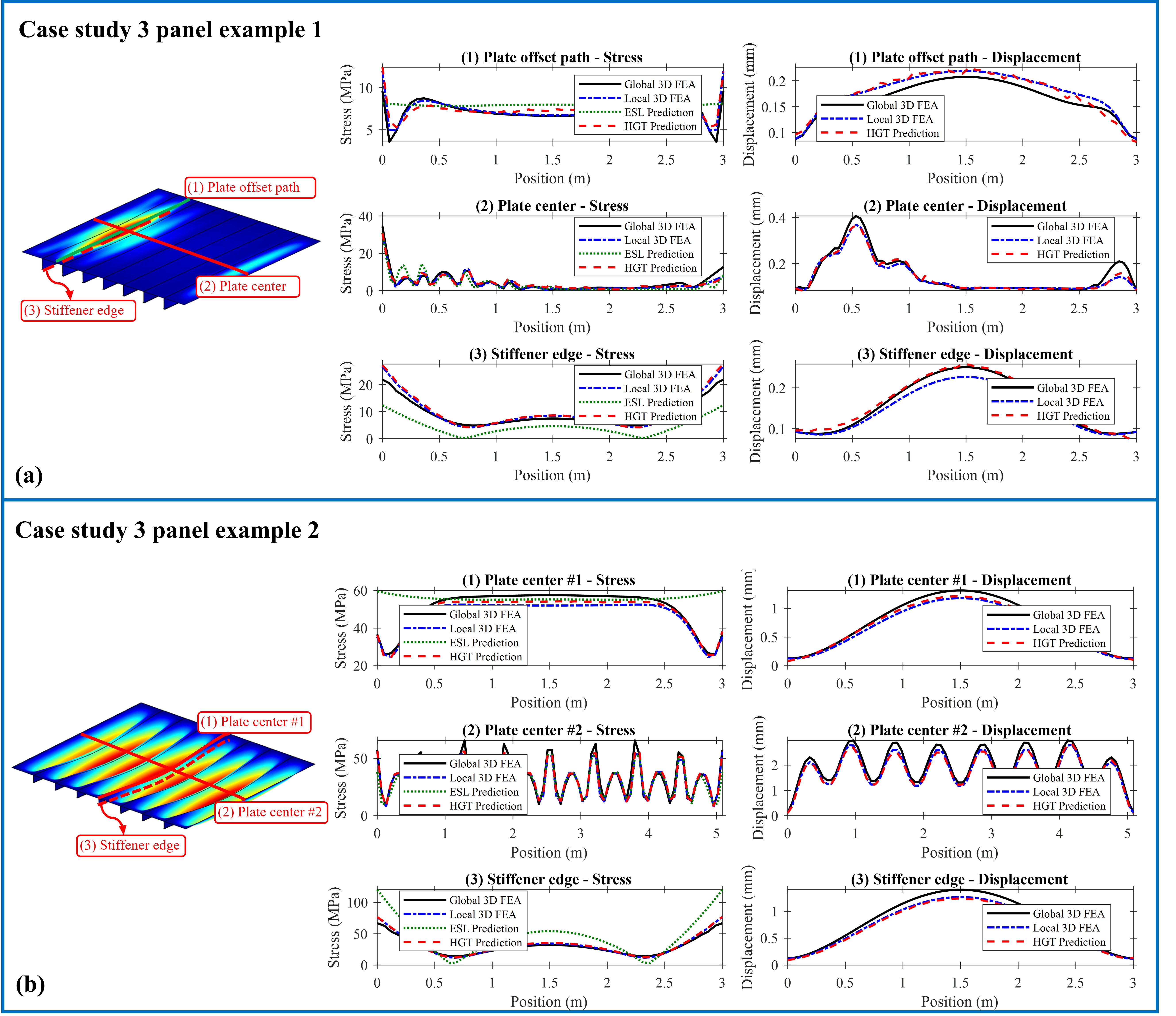}
    \caption{Comparison of the stress and total displacement for two example panels in case study 1: (a) side panel and (b) bottom panel. Results are shown along three prescribed paths.}
    \label{fig: Paper3_case3_panel_comparison}
\end{figure}

This section assesses the performance of the HGT, which is trained on the data from the local 3D FE model. For each case study, two representative panels were selected from the box beam segments identified in Section~\ref{sec4_2} based on the median accuracy of the total framework predictions. Panels presented here are chosen to showcase a variety of structural conditions, including different load conditions and positions within the box beam. For each selected panel, comparisons are presented along three representative paths. These paths were selected to highlight regions with the largest displacements and stresses. The resulting von Mises stress and total displacement plots, along with their path definitions, are shown in Figs.~\ref{fig: Paper3_case1_panel_comparison}--\ref{fig: Paper3_case3_panel_comparison}. These figures present the predictions of the $\text{HGT}$ model, its direct reference model used for training (the local 3D FE model), the overarching reference model, i.e., the global 3D FE model, and the conventional ESL-based stress prediction method.

Fig.~\ref{fig: Paper3_case1_panel_comparison} presents the results for case study 1, comparing predictions for a top panel and a side panel. The top panel is directly subjected to uniform pressure. The HGT surrogate demonstrates good fidelity across all paths, particularly at locations with high stress and displacement. Specifically, the local 3D FE model predicts a peak stress of $140.7\text{ MPa}$ at the stiffener edge; the HGT surrogate accurately captures this with a prediction of $143.8\text{ MPa}$, corresponding to a $97.8\%$ accuracy. The HGT surrogate also successfully predicts the complex, wavy stress profile along the plate center and provides accurate predictions for the displacement fields across all three paths. For the side panel (Fig.~\ref{fig: Paper3_case1_panel_comparison} (b)), which is not directly loaded, the HGT still exhibits good performance. Although the local 3D FEA curves may deviate from the global 3D FEA solutions (as seen in the displacement comparisons for paths on the plate), the HGT consistently aligns well with the local 3D FEA on which it was trained. This observation reinforces the conclusion from Section~\ref{sec4_1} that the local 3D FE model (essentially a sub-model of the global ESL FE model) is the major error source for the overall framework.

The results for case study 2 are shown in Fig.~\ref{fig: Paper3_case2_panel_comparison}. Similar to case study 1, the HGT maintains high accuracy in predicting the maximum values, which is critical in structural design. In panel example 2 (Fig.~\ref{fig: Paper3_case2_panel_comparison} (b)), the maximum stress is observed at the edge that is the simply supported boundary of the box beam, where the HGT prediction exceeds $99\%$ accuracy. However, the HGT shows deviation in locations where the stress or displacement values are relatively small. For instance, along plate edge 2 in the panel example 1, the relative displacement error is $22.17\%$ while the absolute RMSE is only $0.0235\text{ mm}$. This outcome is caused by the RMSE loss function, which inherently emphasizes regions with larger stress and displacement values and can yield inflated relative errors where the ground-truth values are small.

Fig.~\ref{fig: Paper3_case3_panel_comparison} demonstrates the HGT performance for case study 3, which involves more complex structures and spatially varying loadings. Despite this complexity, the HGT successfully captures the overall trends and local extrema in both stress and displacement fields. For example, in panel example 1, which is the side panel subjected to non-uniform pressure, the HGT accurately predicts the multi-peak stress profile along the plate center. In panel example 2 (bottom panel), the HGT maintains high fidelity. Specifically, at the stiffener edge, the local 3D FE model shows a peak stress of $78.55\text{ MPa}$, while the HGT predicts $80.55\text{ MPa}$, achieving $97.5\%$ accuracy at this critical location. Overall, the panel-level accuracies for the HGT in case study 3 exceed $90\%$ for both stress and displacement fields with respect to the local 3D FEA reference. 

\subsubsection{Comparison of HGT with conventional ESL stress prediction method}\label{sec4_3_2}

Stresses in stiffened panels can be approximated using the conventional $\text{ESL}$ stress prediction method introduced in Section~\ref{sec2_1_2}. In this approach, stress at the plate surface exposed to pressure loading is obtained by superimposing the averaged stress field $\sigma_{av}$ and the local bending stress $\sigma_Q$. For panels without direct pressure loading, the local bending term $\sigma_Q$ is omitted, and panel stresses are simply taken as the averaged stress $\sigma_{\text{av}}$. The original ESL formulation does not directly provide stresses on stiffener webs and flanges. In this study, these are approximated by applying the averaged stress equation (Eq.~\ref{Eq: ESL strain}) but replacing the through-thickness coordinate ($z$) with the distance from the homogenized plate surface to the point under evaluation.

Figs.~\ref{fig: Paper3_case1_panel_comparison}--\ref{fig: Paper3_case3_panel_comparison} demonstrate both the strengths and inherent limitations of the $\text{ESL}$ stress prediction method. The best performance is achieved for panels in case study 2 (Fig.~\ref{fig: Paper3_case2_panel_comparison}), where the $\text{ESL}$ method shows a reasonably accurate prediction for stresses on the plate. Since the panels are not directly subjected to lateral pressure, the $\text{ESL}$ prediction relies on the global average stress $\sigma_{\mathrm{av}}$, which is the main stress component the ESL is designed to predict. In case study 1 (Fig.~\ref{fig: Paper3_case1_panel_comparison}), the $\text{ESL}$ results exhibit reasonable agreement with the global $\text{3D}$ $\text{FEA}$ only for stresses located on the plate center, but accuracy notably decreases along the plate edge. This decrease is anticipated because the cylindrical bending assumption is invalid away from the center, as it neglects other curvatures, leading directly to a loss of accuracy in these regions. A similar decrease in accuracy is observed at the stiffener edges for all panels across all case studies. While the overall stress patterns are qualitatively captured for some panels, the stress predictions on the stiffener edges frequently deviate significantly from the global $\text{3D}$ $\text{FEA}$.

\begin{table}[htbp]
  \centering
  \small
  \caption{Panel-wise stress $\text{RMSE}$ (in MPa): $\text{HGT}$ model and conventional $\text{ESL}$ stress prediction method, both benchmarked against the global 3D FE model (reference).}
  \label{tab: recover_vs_esl_global}
  \begin{tabular}{@{}llcc@{}} 
    \toprule
    Case study & Panel example & HGT &  ESL \\
    \midrule
    \multirow{2}{*}{ 1} & \ 1 & 22.58 & 124.9 \\
    & \ 2 & 2.502 & 9.548 \\
    \midrule 
    \multirow{2}{*}{ 2} &  \ 1 & 1.125 & 5.762 \\
    &  \ 2 & 10.79 & 62.16 \\
    \midrule 
    \multirow{2}{*}{ 3} &  \ 1 & 1.048 & 3.590 \\
    &  \ 2 & 2.328 & 13.411 \\
    \bottomrule
  \end{tabular}
\end{table}

To quantitatively evaluate the improvement achieved by the proposed approach, Table~\ref{tab: recover_vs_esl_global} compares the performance of the HGT against the conventional $\text{ESL}$ in terms of stress predictions across the entire panels for the examples detailed in $\text{Section}~\ref{sec4_3_1}$, benchmarked against the global 3D FE model. It is evident from $\text{Table}~\ref{tab: recover_vs_esl_global}$ that the $\text{HGT}$ surrogate consistently demonstrates less error with respect to the global 3D FE model than the $\text{ESL}$ prediction method across all panel examples and case studies. Overall, the proposed method reduces the panel-wise stress $\text{RMSE}$ by at least a factor of three.

While the $\text{ESL}$ method provides a rapid overall prediction of the stress field, its oversimplified assumptions for local stress calculation limit its accuracy, particularly for complex geometries and loading conditions. Additionally, the original $\text{ESL}$ formulation (Section~\ref{sec2_1_2}) treats the plate spans between stiffeners as beams with fixed-fixed boundary conditions and was primarily intended for predicting only the $y$-direction stress component along the center of the plate. This inherent limitation explains the observed errors for off-center paths. A more comprehensive  discussion of the $\text{ESL}$ error along different paths and the effect of other boundary conditions can be found in $\text{Appendix B}$.

In contrast, the data-driven surrogate used in this study does not rely on a priori boundary condition assumptions for stress reconstruction. Its accuracy is determined primarily by the quality of the training data and by the fidelity of the boundary-displacement recovery that drives the local analysis. This makes it a suitable surrogate for the analysis of stiffened panels.

\section{Conclusion and future work}

A hybrid framework is presented for the first time that couples the equivalent single layer (ESL) method with the graph neural network (GNN) to perform global and local analysis of ship hull girders. The global stage employs a coarse mesh ESL model to efficiently obtain the displacement field of the ship hull structure. A boundary DOFs recovery procedure was developed and applied to reconstruct the detailed displacements and rotations along the stiffener web and flange edges, which are essential inputs for the subsequent local analysis. At the local stage, stiffened panels are encoded using an improved heterogeneous graph representation approach. The heterogeneous graph transformer ($\text{HGT}$) model then predicts panel-level stress and displacement fields using the recovered boundary DOFs. The HGT model is trained using the local 3D FE model of a stiffened panel. The framework was validated on three box beam case studies, each featuring distinct geometries and loading conditions. The key findings are summarized as follows:

\begin{itemize}
    \item The proposed framework accurately predicts the stress and displacements of ship hull structures with high fidelity while circumventing the need for computationally expensive, detailed global 3D FEA.
    \item The ESL model and the boundary DOFs recovery procedure generate the dominant portion of the total framework error.
    \item The $\text{HGT}$ surrogate consistently demonstrates high agreement with its training reference, the local 3D FEA model solution.
    \item The $\text{HGT}$ surrogate, driven by recovered boundary DOFs, significantly outperforms the conventional $\text{ESL}$ stress prediction method when predicting local stresses.
\end{itemize}

In summary, the proposed hybrid framework offers a powerful and efficient tool for early-stage and iterative ship structural design. Once the local $\text{HGT}$ surrogate is properly trained, the hull girder's detailed response can be analyzed rapidly by only conducting the computationally inexpensive global $\text{ESL}$ $\text{FEA}$. This capability makes the framework highly suitable for optimization procedures at early design stages, where rapid, high-fidelity hull girder analysis is essential. Although 6000 data samples were prepared for training and testing the HGT surrogate model, a substantially smaller dataset can achieve comparable high-accuracy performance. This efficiency makes the framework practical for applications with limited data. 

Future work could focus on improving the accuracy of global modeling and achieving higher accuracy of the boundary recovery, as these steps control the end-to-end accuracy of the entire framework. Additionally, to reduce the high cost associated with preparing supervised training data for the surrogate, data-efficient learning strategies need to be explored, such as unsupervised physics-informed learning, which can leverage governing equations to minimize the reliance on expensive labeled data.

\section*{Authorship contribution statement}
\textbf{Yuecheng Cai}: Conceptualization, Methodology, Modelling, Validation, Writing - original draft, Writing - review, Writing - editing.

\noindent\textbf{Jasmin Jelovica}: Conceptualization, Methodology, Supervision, Funding acquisition, Writing - review, Writing - editing.

\section*{Acknowledgments}
This research was supported by the Natural Sciences and Engineering Research Council of Canada (NSERC) Discovery Grant [grants IRCPJ 550069-19, RGPIN-2025-04421 and DGDND-2025-04421].


\appendix

\section{ESL stress prediction with different boundary conditions}\label{app: esl_local}
\setcounter{figure}{0}
\setcounter{table}{0}

The ESL method requires the bending stress determined from a beam with specified BCs, to be superimposed onto the global $\text{ESL}$ solution. To assess how BC assumptions influence the accuracy of the conventional $\text{ESL}$ stress prediction method, we compared $\text{ESL}$ stress predictions under three common BC assumptions:
\begin{itemize}
    \item Fixed-Fixed ($\mathrm{FF}$): Both sides fixed.
    \item Simply Supported ($\mathrm{SS}$): Both sides simply supported.
    \item Guided-Fixed ($\mathrm{GF}$): One side guided, the other side fixed.
\end{itemize}

\begin{figure}
    \centering
    \includegraphics[width=0.75\linewidth]{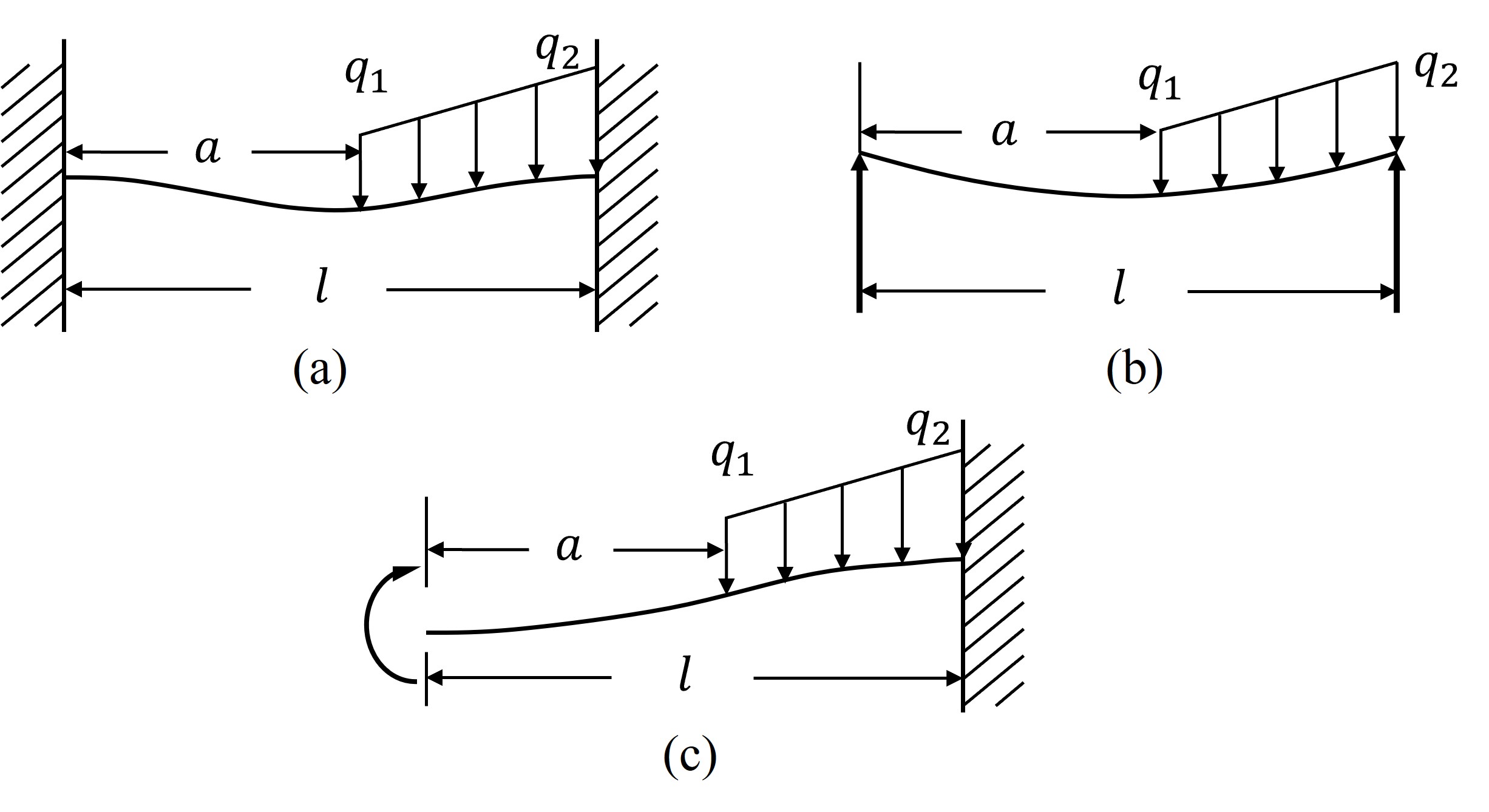}
    \caption{Beam with different boundary conditions subjected to partial distributed load: (a) both sides fixed, (b) both sides simply supported, (c) left side guided, right side fixed.}
    \label{Fig: Paper3_appendix_localization_BCs}
\end{figure}

\begin{figure}[htbp]
    \centering
    \includegraphics[width=0.85\textwidth]{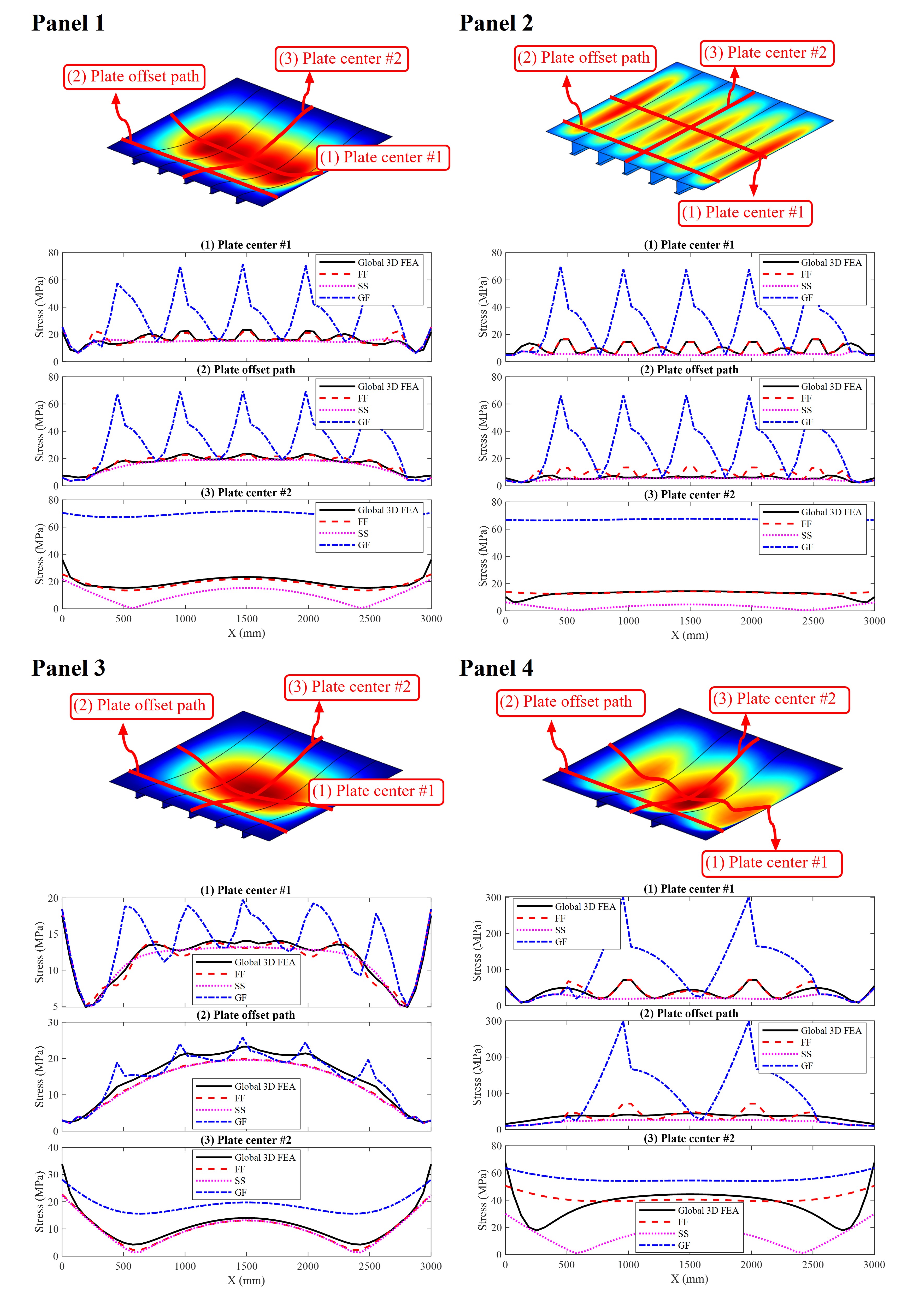}
    \caption{ESL stress predictions with different boundary condition assumptions for four test panels with distinct geometries.}
    \label{fig: Paper3_appendix_ESL_localization}
\end{figure}

The bending moment equations for the $\mathrm{FF}$ case are presented in $\text{Section}~\ref{sec2_1_2}$. The subsequent equations present the bending moments $M(y)$ for the simply supported ($\mathrm{SS}$) and guided-fixed ($\mathrm{GF}$) beams under uniform, trapezoidal, and triangular loadings:

\[
M_{\mathrm{SS},\,\mathrm{uniform}}(y) =
\dfrac{q\,y\,(l - 4y)}{8}
\]

\[
M_{\mathrm{SS},\,\mathrm{trapezoidal}}(y) =
\dfrac{(q_1 - q_2)\,y^{3}}{6\,l} \;-\; \dfrac{q_1\,y^{2}}{2}
\;+\; \left(\dfrac{l\,q_1}{8} - \dfrac{l\,(q_1 - q_2)}{10}\right) y
\]

\[
M_{\mathrm{SS},\,\mathrm{triangular}}(y) =
\begin{cases}
-\dfrac{q_2\,y\,(a - l)^{3}\,(a + 4l)}{40\,l^{3}}, & y \le a,\\[8pt]
-\dfrac{q_2\,(a - y)^{3}}{6\,(a - l)} \;-\; \dfrac{q_2\,y\,(a - l)^{3}\,(a + 4l)}{40\,l^{3}}, & \text{otherwise}.
\end{cases}
\]

\[
M_{\mathrm{GF},\,\mathrm{uniform}}(y)=
\dfrac{q\!\left(l^{2}-3y^{2}\right)}{6}
\]

\[
M_{\mathrm{GF},\,\mathrm{trapezoidal}}(y)=
\dfrac{l^{2}\,q_1}{6}-\dfrac{q_1\,y^{2}}{2}-\dfrac{l^{2}\!\left(q_1-q_2\right)}{24}
+\dfrac{y^{3}\!\left(q_1-q_2\right)}{6\,l}
\]

\[
M_{\mathrm{GF},\,\mathrm{triangular}}(y)=
\begin{cases}
-\dfrac{q_2\,(a-l)^{3}}{24\,l}, & y \le a,\\[8pt]
-\dfrac{q_2\,(a-l)^{3}}{24\,l}\;-\;\dfrac{q_2\,(a-y)^{3}}{6\,(a-l)}, & \text{otherwise.}
\end{cases}
\]

\noindent where $l$ represents the length of the beam (i.e., the width of the plate between stiffeners), $q_{1}$ and $q_{2}$ represent the magnitudes of the load on the left- and right-hand sides of the distribution, respectively, and $a$ defines the starting location of the partially distributed triangular load. The specific BCs and load parameters are illustrated in Fig.~\ref{Fig: Paper3_appendix_localization_BCs}.

\begin{table}[h]
\centering
\small
\caption{Geometric variables for stiffened panels in Fig.~\ref{fig: Paper3_appendix_ESL_localization}.}
\label{tab: ESL compar panel geometry}
\begin{tabular}{@{}lcccc@{}}
\toprule
Category (unit)                 & Panel 1 & Panel 2 & Panel 3 & Panel 4\\ 
\midrule
Plate thickness (mm)    & 10 & 10 & 20 & 10 \\
Web thickness (mm)      & 5 & 20 & 5 & 5 \\
Web height (mm)         & 100 & 200 & 100 & 100 \\
Flange thickness (mm)   & 5 & 20 & 5 & 5 \\
Flange width (mm)       & 50 & 100 & 50 & 50 \\
Number of stiffeners    & 5 & 5 & 5 & 2 \\
\bottomrule
\end{tabular}
\end{table}

To thoroughly analyze the influence of the BCs, we tested four panels (Fig.~\ref{fig: Paper3_appendix_ESL_localization}) with distinct geometric settings chosen to represent a broad range of stiffened panel characteristics. The considered characteristics encompass structural extremes, including the smallest and largest stiffeners, the thinnest and thickest plates, and panels with nearly average or the fewest stiffeners, see Table~\ref{tab: ESL compar panel geometry}. These four panels are located at the top of the middle segment of the single-unit box beam. For each stiffened panel, we compared the predicted stresses along three paths on the plate to assess the performance at different region of the plate. Plate center 1 runs transverse to the stiffeners. This location aligns with the region primarily targeted by the conventional ESL formulation \cite{avi2015equivalent}. We chose two additional paths to test the $\text{ESL}$ beam assumption's effectiveness away from this center. Plate center 2 runs parallel to the stiffeners, covering the panel's entire length, and the plate offset path, located $10\%$ of panel's length from the edge.

The comparisons along the three paths for all four test panels are shown in Fig.~\ref{fig: Paper3_appendix_ESL_localization}. The global 3D FEA solution is included as the reference. It can be observed across all panels that $\text{ESL}$ with the $\mathrm{FF}$ boundary assumption predicts the magnitude and shape of the stress distribution with the highest accuracy. Switching to the other $\text{BCs}$ introduces systematic bias: $\mathrm{GF}$ tends to overestimate and $\mathrm{SS}$ to underestimate the stress over the same paths. It is worth noting that closer to the plate edge (along the plate offset path), the accuracy of all three $\text{BC}$ assumptions decreases, including the $\mathrm{FF}$ case that performs well at the plate center 1. This outcome is consistent with the discussion in $\text{Section}~\ref{sec4_3_2}$, where the cylindrical bending assumptions were shown to become invalid at regions close to the panel boundary. This decrease in accuracy is also visible along the plate center 2, demonstrating inaccuracy of the ESL there even with the FF boundary assumption.  Focusing on plate center 1 and performance under the FF assumption, it is evident that the highest accuracy occurs for panel 2, whose sturdiest stiffeners and thinnest plate geometry allow the FF BC to hold true.

\section{Effect of dataset size on HGT accuracy}\label{app: training_size}
\setcounter{figure}{0}
\setcounter{table}{0}

To examine the effect of training dataset size on model performance, the HGT was systematically trained using datasets of varying sizes, ranging from 100 to 4800 training samples. Each data sample contained stress information of an entire panel. To ensure a fair comparison, every model used identical hyperparameters and architectures.

\begin{figure}[htbp]
    \centering
    \includegraphics[width=0.55\textwidth]{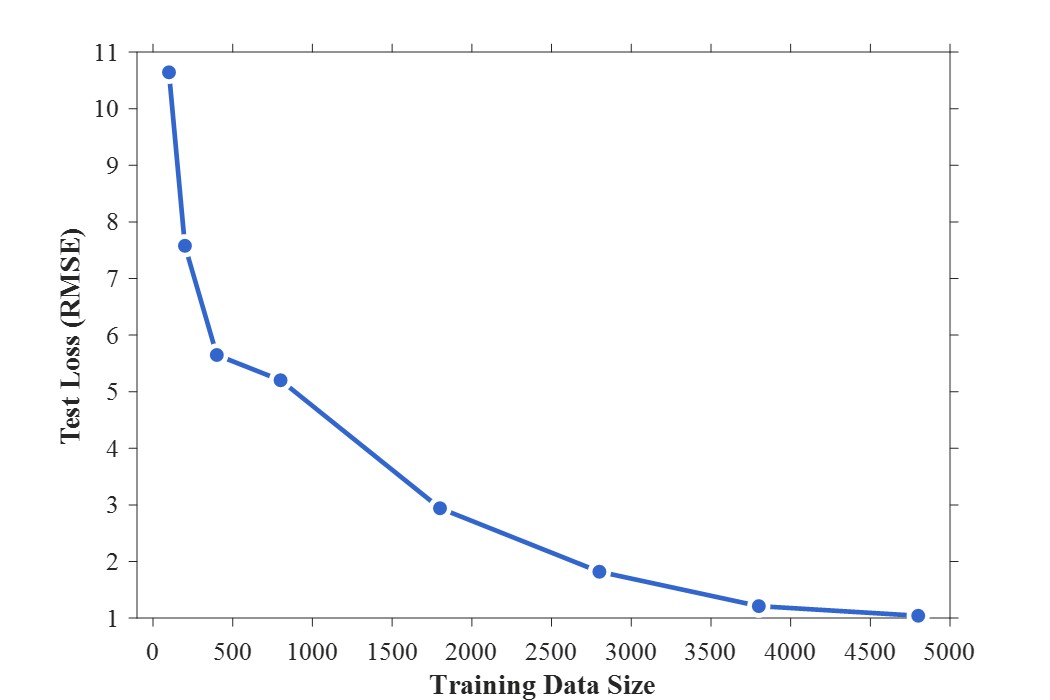}
    \caption{Effect of training dataset size on RMSE performance for von Mises stress prediction.}
    \label{fig: Paper3_appendix_train_size}
\end{figure}

Fig.~\ref{fig: Paper3_appendix_train_size} shows the RMSE of the test set for von Mises stress in Case Study 1 as a function of training dataset size. The curve exhibits three distinct regimes: (1) rapid improvement from 100 to 400 samples, indicating efficient learning of fundamental stress patterns; (2) diminishing returns from 400 to 2800 samples, suggesting the model approaches optimal performance; and (3) near-convergence beyond 3800 samples, where the addition of further data yields marginal benefit to accuracy.

\begin{figure}[htbp]
    \centering
    \includegraphics[width=0.9\textwidth]{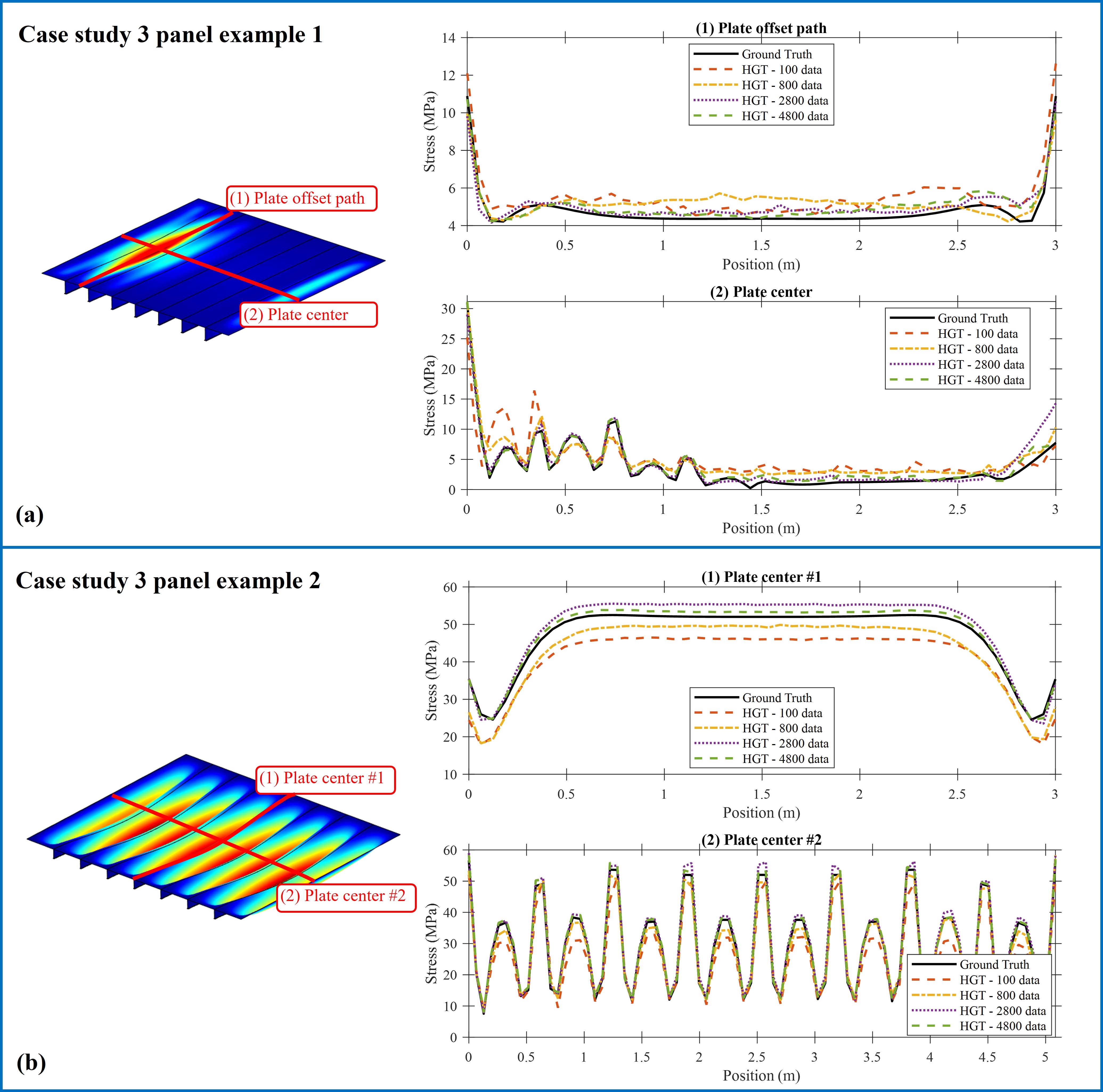}
    \caption{Effect of training dataset size on $\text{HGT}$ performance for von Mises stress prediction in case study 3 panel examples.}
    \label{fig: Paper3_appendix_train_size_path}
\end{figure}

Fig.~\ref{fig: Paper3_appendix_train_size_path} compares stress predictions from models trained with different amounts of data against local 3D FEA results along two selected paths for both panel examples in case study 3. This case study was chosen for visualization due to its complexity. The results clearly demonstrate a progressive improvement in prediction accuracy as the training data increases. Specifically, the model trained with 100 samples shows the greatest deviations from the reference, which is improved as the amount of training data progressively increases to 4800. In this article, a training data size of 4800 panels (6000 with added validation and testing samples) is utilized to ensure the highest possible accuracy of the $\text{HGT}$ model. For practical purposes, however, substantially smaller dataset could suffice.

\end{document}